\documentclass[fleqn,usenatbib]{mnras}

\usepackage{newtxtext,newtxmath}

\usepackage[T1]{fontenc}

\DeclareRobustCommand{\VAN}[3]{#2}
\let\VANthebibliography\thebibliography
\def\thebibliography{\DeclareRobustCommand{\VAN}[3]{##3}\VANthebibliography}

\usepackage{amsmath}
\usepackage{amssymb}
\usepackage{gensymb}
\usepackage{amsfonts}
\usepackage{verbatim}
\usepackage{scalefnt}
\usepackage[percent]{overpic}
\usepackage[T1]{fontenc} 
\usepackage{aecompl} 
\usepackage{ulem}
\usepackage{cleveref}
\usepackage{todonotes}
\usepackage{tikz}
\usepackage{tikz-3dplot}
\usetikzlibrary{angles, quotes}
\usepackage{tkz-euclide}
\usetkzobj{all}
\usepackage{xcolor}
\usepackage{subcaption}
\usepackage{multicol}
\usepackage{float}
\usepackage{mathtools, cuted}

\newcommand{\Nt}{N_t}

\usepackage[switch]{lineno}

\makeatletter
\@namedef{Changes@AuthorColor}{red}
\colorlet{Changes@Color}{red}
\makeatother

\title[Towards a GW Polarisation Prediction]{Towards a Polarisation Prediction for LISA \\
via Intensity Interferometry}

\author[]{{Sandra Baumgartner}$^{1}$,
{Mauro Bernardini}$^{1}$,
{Jos\'{e} R. Canivete Cuissa}$^{1,2}$,
\newauthor
{Hugues de Laroussilhe}$^{1}$,
{Alison M. W. Mitchell}$^{3}$,
{Benno A. Neuenschwander}$^{1}$,
\newauthor
Prasenjit Saha$^{3}$,
{Timoth\'{e}e Schaeffer}$^{1}$,
{Deniz Soyuer}$^{1}$ and {Lorenz Zwick}$^{1}$
\\
$^{1}$Center for Theoretical Astrophysics and Cosmology, Institute for Computational Science, University of Zurich, \\Winterthurerstrasse 190, CH-8057 Z{\"u}rich, Switzerland\\
$^{2}$Istituto Ricerche Solari Locarno (IRSOL),
	Via Patocchi 57, CH-6605 Locarno-Monti, Switzerland\\
$^{3}$Physik-Institut, Universit{\"a}t Z{\"u}rich, Winterthurerstrasse 190, CH-8057 Z{\"u}rich, Switzerland\\
}

\date{Accepted 2020 August 26. Received 2020 August 24; in original form 2020 July 29}

\pubyear{2020}

\begin{document}
\label{firstpage}
\pagerange{\pageref{firstpage}--\pageref{lastpage}}
\maketitle

\begin{abstract}
 Compact Galactic binary systems with orbital periods of a few hours are expected to be detected in gravitational waves (GW) by LISA or a similar mission. At present, these so-called verification binaries provide predictions for GW frequency
 and amplitude.  A full polarisation prediction would provide a new method to calibrate LISA and other GW observatories, but requires resolving the orientation of the binary on the sky, which is not currently possible.  We suggest a method to determine the elusive binary orientation and hence predict the GW polarisation, using km-scale optical intensity interferometry. The most promising candidate is CD--30$^\circ$ 11223, consisting of a hot helium subdwarf with $m_B=12$ and a much fainter white dwarf companion, in a nearly edge-on orbit with period 70.5~min.  We estimate that the brighter star is tidally stretched by 6\%. Resolving the tidal stretching would provide the binary orientation.  The resolution needed is far beyond any current instrument, but not beyond current technology.  We consider scenarios where an array of telescopes with km-scale baselines and/or the Very Large Telescope (VLT) and Extremely Large Telescope (ELT) are equipped with recently-developed kilo-pixel sub-ns single-photon counters and used for intensity interferometry.  We estimate that a team-up of the VLT and ELT could measure the orientation to $\pm1^\circ$ at 2$\sigma$ confidence in 24 hours of observation.
\end{abstract}

\begin{keywords}
gravitational waves -- 
techniques: interferometric --
stars: individual: CD$-30^\circ$ 11223
\end{keywords}



\section{Introduction}\label{sec:introduction}

Gravitational-wave detections are so far all transient events with no
advance warning.  This will change when laser interferometers in space
make lower-frequency gravitational waves detectable, because some
Galactic binaries (known as LISA verification binaries) are predicted
to be detectable via gravitational waves \citep{2006CQGra..23S.809S}.
It would be interesting to have predictions, not only for the expected
strain, but for the expected polarisation as well.  Gravitational-wave
polarisation is now measurable \citep[see
  e.g.,][]{PhysRevLett.119.141101} since the VIRGO detector has a
different orientation from the two LIGO detectors.  In the future, it
will be possible to test for additional polarisation modes \citep[see
  e.g.,][]{2017PhRvD..96d2001I,2018PhRvD..98d4025P}, which if they
exist, would imply new physics.

The gravitational wave polarisation of a LISA verification binary
could be predicted if the orientation of the system in the sky (inclination and position angle)
were somehow measured from its electromagnetic radiation.
This is where intensity interferometry comes into play. 
First introduced by \cite{hanbo}, it was used to measure sizes of stars by counting coincident photons from an object with a pair of telescopes. By varying the separation between the telescopes, they showed that it is possible to measure the spatial correlation function of the source brightness distribution. However, with the detectors available at the time, intensity interferometry was only feasible for some of the brightest stars, with B magnitude $< 2.5$ \citep{Hanburyetal74}.
With a new generation of photon detectors now available, intensity
interferometry has been revived in recent years
\citep{2016SPIE.9907E..0NZ,2018SPIE10701E..0XW,2018JMOp...65.1336M,
  2018MNRAS.480..245G, 2020MNRAS.491.1540A, Abeysekara2020}.
Especially interesting is the proposal to use the Cherenkov Telescope
Array (CTA) for intensity interferometry \citep{Dravins2012}, which
would make baselines of up to 2\thinspace km possible.

In this paper we propose a ``multi-messenger'' method to predict the gravitational-wave polarisation that would be observed by LISA when probing a binary system, and in particular the system CD$-30^\circ$ 11223, which is the optically
brightest and most widely spaced of the LISA verification binaries \citep{kupfer}. The system consists of a white dwarf and a hot helium subdwarf.  The former is the more massive of the two objects, but contributes less than 1\% of the light.  As a result, resolving the binary as two stars through interferometry is not a prospect, but resolving the tidal stretching of the brighter star is more promising. We suggest that, with the use of intensity interferometry, one can infer the shape of the tidally stretched subdwarf on the sky, thus determining the orientation of the system with respect to the Earth.  

The paper is structured as follows: In Section \ref{sec:methods} we present our method for predicting the polarisation via intensity interferometry by exploiting the tidal effects in a binary system. In Section \ref{sec:binary} we give the system parameters of CD$-30^\circ$ 11223 and calculate the tidal deformation of the subdwarf. We present our results for various combinations of telescope arrays used for intensity interferometry. We discuss our findings and present our concluding remarks in Section \ref{sec:discussion} .

\section{Methods}
\label{sec:methods}
\subsection{The missing ingredient}\label{subsec:methodsGW}
The LISA mission will consist of three satellites in orbit around the Sun. Photons exchanged between the satellites carry information about their relative distances. If a GW passes through the Solar System, the photons' paths will be stretched and shrunk accordingly. This effect enables LISA to detect the passing GW. Considerable effort has been put into creating larger and more precise catalogues of expected GW signals \citep[see e.g.][ for different approaches]{templ1,templ2,templ3,templ4}. 
Such signals \citep[see e.g.][]{lisa1,lisa2} originate from various types of sources (stellar mass binaries, super-massive binaries, extreme mass ratio inspirals) located in very different astrophysical settings. Large catalogues are necessary because LISA data analysis primarily works by means of parameter fitting. 
 Roughly described, it will detect a GW signal and then try to match it to a GW template from a catalogue \citep[see e.g.,][for information on matched filtering]{filtering}. One can then trace back the properties of the source by noting what templates best fit the signal. Even though there is much talk about the capacity of LISA to test the strongest of gravitational regimes \citep[see e.g.][]{nohair}, it is clear that results obtained by parameter fitting must be subject to careful analysis as the process is liable to various kinds of degeneracy.
Indeed, very different GW sources might produce signals that happen to be indistinguishable by parameter estimation if they are located and oriented in an inconvenient manner. Further degeneracies arise by including alternative theories of gravity, where the signal from a source in one theory might be confused with a signal from another source in a different theory. Furthermore, it is possible that some unexpected systematic error in the experimental setup would lead to systematic errors in parameter estimation, with little possibility of correction.

To get around the process of parameter fitting, it is necessary to produce a \textit{unique} prediction for a gravitational wave signal. In other words, one must identify a promising source of gravitational waves, resolve its physical parameters with independent measurements, assume a theory of gravity and predict both the strain and the polarisation amplitudes LISA would measure from it. Only such a prediction would, in principle, allow for an independent test of General Relativity. More realistically, due to uncertainties in the measurements and unknown systematics, it could be used as a calibration/validation for LISA and similar GW detectors. 

As noted before, there are no sources of GW that have been completely resolved by optical methods. Even in the case of the most widely spaced LISA verification binary CD$-30^\circ$ 11223, the angular size is far below even the milli-arcsecond scale. In particular, even though it is known that its orbital plane is almost edge on, the alignment of the plane is completely undetermined. Crucially, this remaining orientation angle controls the relative amplitude of the predicted gravitational wave polarisations. The objective of this paper is therefore set:  we wish to present a method to resolve the orientation of luminous GW sources in the sky by means of intensity interferometry, thus determining the remaining physical parameter of CD$-30^\circ$ 11223 that is needed for a complete GW prediction.

 \begin{figure}
    \centering
    \includegraphics[width=\columnwidth]{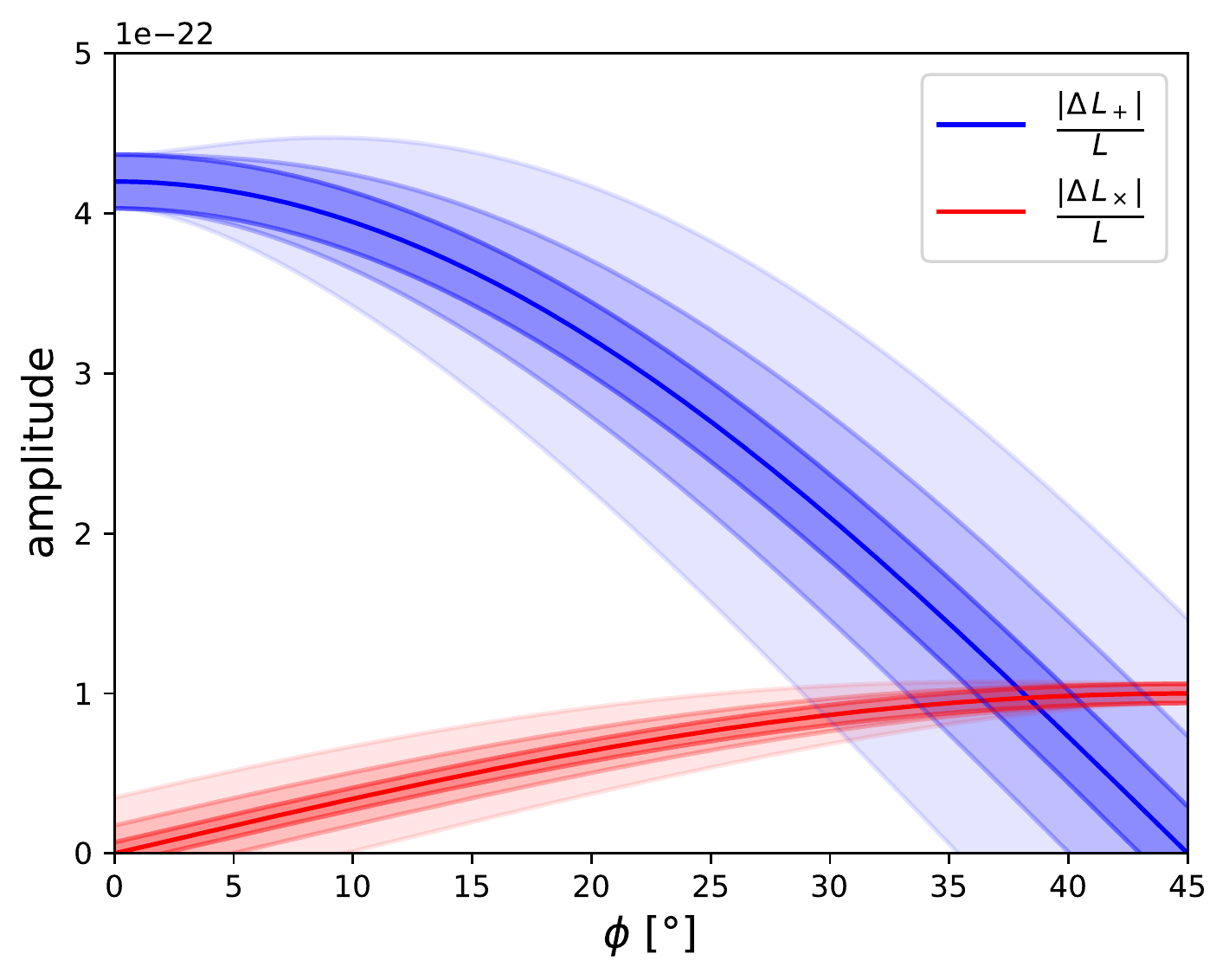}
    \caption{The full range of possible polarisation amplitudes  $\frac{\Delta L_+}{L}$ and $\frac{\Delta L_\times}{L}$ for the binary system CD$-30^\circ$ 11223 as a function of the unresolved orientation angle $\phi$. The uncertainties $\sigma_+$ and $\sigma_\times$ on the prediction are represented by the shaded areas. For this plot we have chosen three representative uncertainties in the angle $\phi$ of $\Delta_{\phi}=\left\{2^\circ, 5^\circ,10^\circ\right\}$. The lightest shades of blue and red respectively correspond to an uncertainty of $10^\circ$. The figure only shows a range between $0^\circ$ and $45^\circ$ as the  amplitudes and their errors repeat periodically in $2\phi$.
    } 
    \label{fig:GW_error_plot_amplitudes}
\end{figure}

\begin{figure}
\resizebox{0.5\textwidth}{!}{
\begin{subfigure}{0.4\textwidth}
\begin{tikzpicture} 
\colorlet{Colorlightblue}{blue!10}
\colorlet{Colorlightorange}{orange!10}
\colorlet{Colorlightbrown}{brown!10}
\coordinate (O) at (0,0,0);
\coordinate (X) at (0,0,3);
\coordinate (Y) at (3,0,0);
\coordinate (Z) at (0,3,0);
\coordinate (L) at (0.7,1.5,0);
\coordinate (Ly) at (0.7,0,0);
\coordinate (Lz) at (0,1.5,0);
\coordinate (P) at (-0.3,2.9,0.3);
\coordinate (Q) at (0.3,2.5,-0.3);
\coordinate (M) at (0,2.7,0);
\coordinate (Ba) at (-1.5,0.7,0);
\coordinate (Bb) at (1.5,-0.7,0);
\draw[thick,-, color=red] (Q) - -(M);
\begin{scope}
\clip(0,0,-3)--(0,0,0)--(3,-1.4,0)--cycle;
\filldraw[color=brown, opacity=0.5, rotate=-25.02] (0,0,0) circle [x radius=1cm, y radius=5mm];
\end{scope}
\draw[color=brown, fill=brown] (0.8, -0.3, -0.25) circle [x radius=1.2mm, y radius=1.2mm, rotate=0];
\draw[color=Colorlightorange,fill=Colorlightorange, opacity=0.7] (-1,0,0) -- (3,0,0) -- (3,-1,0) -- (-1,-1,0) -- cycle;
\draw[color=Colorlightblue,fill=Colorlightblue, opacity=0.7] (-1,0,-1) -- (3,0,-1) -- (3,0,0) -- (-1,0,0) -- cycle;
\begin{scope}
\clip(0,0,-3)--(0,0,0)--(-3,1.4,0)--cycle;
\filldraw[color=brown, opacity=0.5, rotate=-25.02] (0,0,0) circle [x radius=1cm, y radius=5mm];
\end{scope}
\draw[color=Colorlightorange,fill=Colorlightorange, opacity=0.7] (-1,0,0) -- (Y) -- (3,3,0) -- (-1,3,0) -- cycle;
\begin{scope}
\clip(0,0,3)--(0,0,0)--(3,-1.4,0)--cycle;
\filldraw[color=brown, opacity=0.5, rotate=-25.02] (0,0,0) circle [x radius=1cm, y radius=5mm];
\end{scope}
\draw[color=Colorlightblue,fill=Colorlightblue, opacity=0.7] (-1,0,0) -- (3,0,0) -- (3,0,3) -- (-1,0,3) -- cycle;
\begin{scope}
\clip(0,0,3)--(0,0,0)--(-3,1.4,0)--cycle;
\filldraw[color=brown, opacity=0.5, rotate=-25.02] (0,0,0) circle [x radius=1cm, y radius=5mm];
\end{scope}
\draw[color=brown, fill=brown] (-0.8, 0.3, 0.25) circle [x radius=0.7mm, y radius=0.7mm, rotate=0];
\tkzMarkRightAngle[draw=brown,size=.15](L,O,Ba);
\draw[thick,->] (O) - -(X);
\draw (X) node[anchor=north east, left]{$\mathbf{\hat{x}}$};
\draw[thick,->] (O) - -(Y);
\draw (Y) node[anchor=north east, right]{$\mathbf{\hat{y}}$};
\draw[thick,->] (O) - -(Z);
\draw (Z) node[anchor=west, above]{$\mathbf{\hat{z}}$};
\draw[thick,->, color=brown] (O) - -(L);
\draw (L) node[anchor=south, color=brown]{$\qquad\qquad\quad\mathbf{L}=(0, L_y, L_z)$};
\draw[dotted,-, color=brown] (L) - -(Ly);
\draw[dotted,-, color=brown] (L) - -(Lz);
\draw[thick,-, color=red] (P) - -(M);
\pic [draw, <-, color=black,
      angle radius=8mm, angle eccentricity=1.3,
      "$\mathcal{I}$"] {angle = L--O--Z};
\draw (Q) node[anchor=west, right, color=red] {LISA};
\end{tikzpicture}
\subcaption{ }
\label{subfig:GW_inclination}
\end{subfigure}
}
\resizebox{0.5\textwidth}{!}{
\begin{subfigure}{0.4\textwidth}
\hspace{0.8cm}
\begin{tikzpicture}[baseline={([yshift=5pt]current bounding box.north)}]
\colorlet{Colorlightblue}{blue!10}
\colorlet{Colorlightorange}{orange!10}
\coordinate (O) at (0,0,0);
\coordinate (X) at (0,0,3);
\coordinate (Y) at (3,0,0);
\coordinate (Z) at (0,3,0);
\coordinate (P) at (-0.8,2.5,1.1);
\coordinate (Pxy) at (-0.8,0,1.1);
\coordinate (Pyz) at (-0.8,2.5,0);
\coordinate (Q) at (0.8,0.5,-1.1);
\coordinate (Qxy) at (0.8,0,-1.1);
\coordinate (Qyz) at (0.8,0.5,0);
\coordinate (M) at (0,1.5,0);
\coordinate (Ym) at (-1,0,0);
\coordinate (Zm) at (0,-1,0);
\coordinate (Yn) at (-0.8,1.5,0);
\coordinate (Yp) at (0.8,1.5,0);
\coordinate (pl1) at (-0.8,1.5,-0.5);
\coordinate (pl2) at (0.8,1.5,-0.5);
\coordinate (pl3) at (0.8,1.5,0.5);
\coordinate (pl4) at (-0.8,1.5,0.5);
\draw[color=Colorlightblue,fill=Colorlightblue, opacity=0.8] (pl1) -- (pl2) -- (Yp) -- (Yn) -- cycle;
\draw[thick,-, color=blue] (Qxy) - -(O);
\draw[thick,-, line width=0.5mm, color=red] (Q) - -(M);
\draw[color=Colorlightorange,fill=Colorlightorange, opacity=0.7] (-1,0,0) -- (3,0,0) -- (3,-1,0) -- (-1,-1,0) -- cycle;
\draw[color=Colorlightblue,fill=Colorlightblue, opacity=0.7] (-1,0,-1) -- (3,0,-1) -- (3,0,3) -- (-1,0,3) -- cycle;
\pic [draw, ->, color=blue,
      angle radius=8mm, angle eccentricity=1.3,
      "$\phi$"] {angle = Y--O--Qxy};
\draw[color=Colorlightorange,fill=Colorlightorange, opacity=0.7] (-1,0,0) -- (Y) -- (3,3,0) -- (-1,3,0) -- cycle;
\draw[thick,-, color=orange] (Qyz) - -(M);
\pic [draw, <-, color=orange,
      angle radius=7mm, angle eccentricity=1.6] {angle = Qyz--M--Yp};
      \draw[color=Colorlightblue,fill=Colorlightblue, opacity=0.8] (Yn) -- (Yp) -- (pl3) -- (pl4) -- cycle;
\draw[thick,->] (O) - -(X);
\draw (X) node[anchor=north east, left]{$\mathbf{\hat{x}}$};
\draw[thick,->] (O) - -(Y);
\draw (Y) node[anchor=north east, right]{$\mathbf{\hat{y}}$};
\draw[thick,->] (O) - -(Z);
\draw (Z) node[anchor=west, left]{$\mathbf{\hat{z}}$};
\draw[dotted,->, color=black] (P) - -(Pxy);
\draw[dotted,->, color=black] (P) - -(Pyz);
\draw[thick,-, color=blue] (O) - -(Pxy);
\draw[thick,-, color=orange] (M) - -(Pyz);
\pic [draw, ->, color=blue,
      angle radius=8mm, angle eccentricity=1.3,
      "$\phi$"] {angle = Ym--O--Pxy};
\pic [draw, <-, color=orange,
      angle radius=7mm, angle eccentricity=0.85] {angle = Pyz--M--Yn};
\draw[thick,-, line width=0.5mm,color=red] (M) - -(P);
\draw (1.3,0.8) node[anchor=west, right, color=red] {LISA};
\draw (0.8,0.7,-0.6) node[anchor=east, left, color=orange] {$\psi$};
\draw (-0.7,1.9,-0.55) node[anchor=east, left, color=orange] {$\psi$};
\draw (-0.2,-0.5) node[anchor=west, right, color=blue] {$\mathbf{\hat{x}}\mathbf{\hat{y}}$ projection};
\draw (0.1,2.5) node[anchor=west, right, color=orange] {$\mathbf{\hat{y}}\mathbf{\hat{z}}$ projection};
\end{tikzpicture}
\subcaption{ }
\label{subfig:GW_LISAangles}
\end{subfigure}
}

\caption{Illustration of the coordinate system for which the centre of mass of the binary system serves as the origin. 
The $z$-axis points towards LISA, thereby determining the propagation direction of the gravitational waves.
The $y$-axis is chosen such that the angular momentum $\mathbf{L}$ of the binary system lies in the $y$-$z$ plane. The angle between $\mathbf{L}$ and the $z$-axis is the inclination $\mathcal{I}$.
Figure \ref{subfig:GW_LISAangles} shows the definition of the two angles $\psi$ and $\phi$ related to LISA (red line). The blue and orange lines are the projections of LISA onto the $x$-$y$ plane and the $y$-$z$ plane respectively.}
\label{fig:GW_angles_definition}
\end{figure}
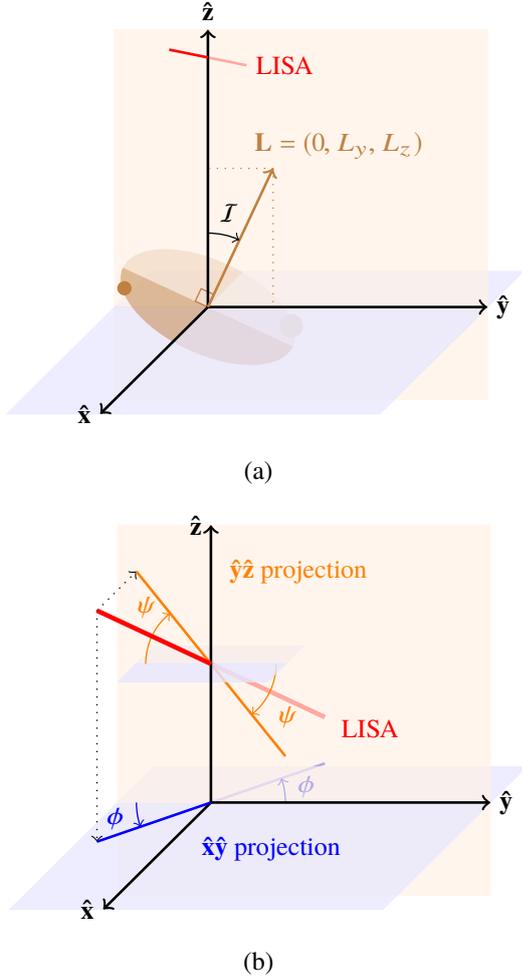

\subsubsection{Coordinate System}
This subsection is dedicated to the construction of a convenient coordinate system incorporating both a binary and a GW detector. For the sake of simplicity, we model the GW detector as consisting of only one arm (generalising to more arms simply means keeping track of more angles). We align the $z$-axis of the coordinate system with the "line of sight direction" from the binary's centre of mass to the midpoint of our idealised detector arm. This choice fixes the $x$-$y$ or ``transverse'' plane in which GWs will produce a measurable strain. The angular momentum vector of the binary system will in general be inclined with respect to the $z$-axis. We define the angle $\mathcal{I}$ as this inclination. We then align the $y$-axis with the projection of the orbit's angular momentum in the transverse plane. This choice of coordinates also fixes the orientation of the detector arm. It will be inclined with respect to the $z$-axis with an angle $\psi$ and rotated from the $y$-axis with an angle $\phi$. For a real space-borne detector, these last two angles will vary as the spacecraft orbits the sun. 

The chosen coordinate system is illustrated in Figure \ref{fig:GW_angles_definition}. In these coordinates, it is the angle $\phi$ that determines the polarisation amplitudes and is the last missing ingredient for a polarisation prediction of the binary CD$-30^\circ$ 11223. In Figure \ref{fig:GW_error_plot_amplitudes} we show how the predicted polarisation amplitudes of CD$-30^\circ$ 11223 change for different values of the angle $\phi$ and different representative uncertainties $\Delta_\phi$ (see Appendix~\ref{appendix:GW} for a derivation).

\subsection{Tidal Stretching}\label{subsec:tidal_stretching}
In order to determine the angle $\phi$ we need to devise a strategy to characterise the orbital plane of the binary system. One possible solution is given by modelling the tidal forces acting upon the stars. 
The total effect of gravitational interactions paired with large orbital velocities can induce stretches in the orbiting bodies due to extreme tidal forces. These deformations, if observed, can give precise information about the orbital plane, and consequently the angle $\phi$.

In the following paragraphs we present two distinct approaches to estimate the induced deformation.  

\subsubsection{Roche potential}

The Roche potential offers an analytic description for the gravitational potential of two tidally locked, corotating bodies on circular orbits with common orbital period $\omega$. 
If the masses $M_1$, $M_2$ and the separation $a$ of the binary stars are known, the Roche potential takes the form:

\begin{equation}
\Phi(\textbf{r}) = -\frac{GM_{1}}{\left|\textbf{r}-\textbf{r}_{1}\right|} - \frac{GM_{2}}{\left|\textbf{r}-\textbf{r}_{2}\right|} -\frac{1}{2}\omega \left|\textbf{r}\right|^{2}
\end{equation}
where $\omega^{2}=G(M_{1}+M_{2})/a^{3}$ is the orbital frequency and $\textbf{r}_1$ and $\textbf{r}_2$ denote the positions of the stars. 
For each one of the binary objects, the tidal deformation redistributes the mass such that on the surface the potential is constant. This specific enclosing equipotential line exactly describes the deformed shape of the star. We solve the problem of finding this specific value numerically by scanning the range of equipotential lines. The desired potential has to fulfill the condition that the enclosed mass corresponds to that of the undeformed case.
If we neglect other sources of deformation (e.g. fast rotation, oscillations), the
general shape of the star will follow a teardrop Roche lobe. 
We aim to measure the ratio $\kappa$ between the major and minor axis of the object, which we define as $r_{\rm max}$ and $r_{\rm min}$ respectively. This can be easily computed once the deformation of the object is known. In most practical cases we can approximate the shape to a spheroid.

\subsubsection{Variations in radiant intensity in the direction of the observer}
Another approach to determine the deformation caused by tidal stretching is to study variations in the radiant intensity in the direction of the Earth. As the visible area of the star changes during one orbital period due to its deformed shape, the received light on Earth changes accordingly. The ratio between the major and minor axis can then be computed by comparing the minimal and maximal value of this flux. 

Let us assume a spheroid shape for the deformed star and model it as a perfect black body. If the two orbiting bodies are aligned with the line of sight of the observer, the area of the star visible from the Earth is $A_{\mathrm{min}} = \pi r_{\mathrm{min}}^2$ and therefore the radiant intensity in the direction of the Earth is proportional to  $I_{e, \Omega}^{\mathrm{min}} \propto \sigma A_{\mathrm{min}} T_{\mathrm{eff}}^{4}$, where $\sigma$ and $T_{\mathrm{eff}}$ denote the Stefan-Boltzmann constant and the effective stellar temperature. Conversely, if the two stars are placed in a plane perpendicular to the line of sight, then the observed shape corresponds approximately to an ellipse of area $A_{\mathrm{max}} = \pi r_{\mathrm{min}}r_{\mathrm{max}}$, and  the radiant intensity in the direction of the Earth is proportional  to $I_{e, \Omega}^{\mathrm{max}}  \propto \sigma A_{\mathrm{max}} T_{\mathrm{eff}}^{4}$. 
Given that the distance to the sub-dwarf is practically constant, we link the observed flux variations directly to the ratio between the axes $r_{\mathrm{min}}$ and $r_{\mathrm{max}}$:
\begin{equation}
\frac{I_{e, \Omega}^{\mathrm{min}}}{I_{e, \Omega}^{\mathrm{max}}} = \frac{A_{\mathrm{min}}}{A_{\mathrm{max}}} = \frac{\pi r_{\mathrm{min}}^2}{\pi r_{\mathrm{min}}r_{\mathrm{max}}} = \frac{r_{\mathrm{min}}}{r_{\mathrm{max}}} = \kappa \,.
\end{equation}

We discuss the case of the CD$-30^\circ$ 11223 binary system in Sect.\,\ref{sec:binary} below.

These two methods provide independent ways of estimating the aspect ratio between $r_{\mathrm{min}}$ and $r_{\mathrm{max}}$, as the first relies on theoretical background and the second is based on observational data. This allows one to cross-check the results. However, uncertainties arise in both procedures as simplifications are made. The fact that this ratio will often be close to 1 gives rise to significant observational challenges, which we hope to surmount with the aid of intensity interferometry.
A good estimate of this aspect ratio is essential for measuring the orientation of the binary in space, which in turn is needed for a polarisation prediction of the gravitational waves.

\subsection{Intensity interferometry}
\label{sec:inten}
To appreciate the challenge
of determining the orientation of the orbital plane, consider first a fictitious binary consisting of two
touching Sun-like stars in a circular orbit ($M_1=M_2=M_\odot,
a=2R_\odot$) observed from a distance of $D=50\rm\,pc$.  The angular
size and the gravitational-wave frequency and strain of a binary are:
\begin{equation}
\theta \sim \frac aD \qquad
f \propto \left(\frac{a^3}{M_1+M_2}\right)^{1/2} \qquad
h \propto \frac{M_1 M_2}{aD}
\end{equation}
respectively.  For this fictitious system, the values come to
\begin{equation}
\theta \sim 10^{-9}{\rm\,radians} \qquad
f \propto 10^{-4}{\rm\,Hz} \qquad
h \propto 10^{-22}.
\end{equation}
Current optical interferometry of binary stars
\citep[e.g.,][]{2019AJ....157..140L} achieves a resolution of
$0.2\rm\,mas\approx10^{-9}\,radians$, so this imaginary system would
be borderline resolvable.  In gravitational waves, however, it would
be too weak/slow to detect with any planned instrument. For a mission
lifetime $T_{\rm obs}$ of a few years, the characteristic strain
$h\sqrt{fT_{\rm obs}}\sim10^{-20}$.  LISA and TianQin are expected to
reach this level of sensitivity, but only at higher frequencies --- see
Figure 1 in \cite{2019CQGra..36j5011R} for LISA and Figure 2 in
\cite{2018CQGra..35i5008H} for TianQin. In comparison with this
imaginary system, real LISA verification binaries are an order of
magnitude closer in orbital separation and an order of magnitude
further away from us.  This makes them much too small on the sky for
current optical interferometry to resolve.
\begin{figure*}
    \centering
    \includegraphics[width = 0.95\textwidth]{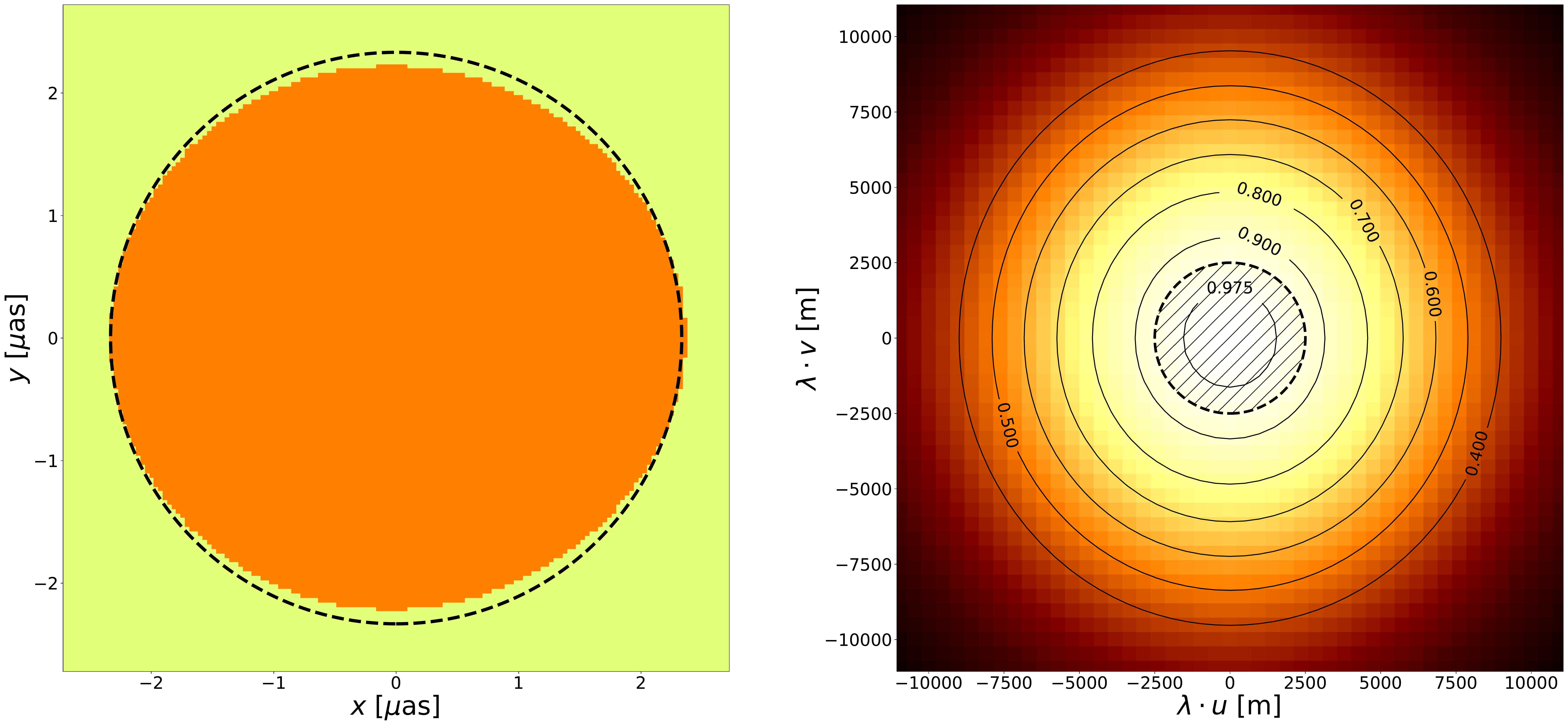}
    \caption{\textbf{Left:} Simulated noise-free image of a uniformly emitting extended source at $\lambda = 350\,$nm in orange. Note that the source is teardrop shaped, same as that calculated for CD--30$^\circ$ 11223 in the next sections. The dashed black curve shows the shape the same object would have in the absence of tidal disruptions.  \textbf{Right:} The Fourier magnitude $|\gamma_{12}|^2$ of the source in the $(u,v)$ plane. Contours show how the magnitude decreases with distance from the center. Not shown in these scales is the full teardrop shaped Airy pattern, which extends beyond 14\thinspace km, having visible local minima and maxima. The dashed area indicates the coverage of the baselines by CTA-S.}
    \label{fig:uv_plane}
\end{figure*}
A possible way to achieve much higher resolution is by means of \textit{intensity} interferometry.  This technique was developed in the 1960s \citep[for the historical development, see][]{hanbury}.  The
detector technology of the time limited its applicability to about 30
blue stars, but the sources that could be observed were resolved to
$\sim0.5\rm\,mas$.

The main idea behind intensity interferometry involves \textit{temporally} correlating the light signals received by a pair of telescopes, separated by a baseline. The measured intensities in both telescopes $\langle I_1\rangle$ and $\langle I_2 \rangle$ (which are averaged over a timescale, characteristic of the setup; the resolution time) will have a cross correlation profile  $\langle I_1 \cdot I_2 \rangle (\boldsymbol{r}_1 - \boldsymbol{r}_2)$ dependent on the projected baseline $\boldsymbol{B} = \boldsymbol{r}_1 - \boldsymbol{r}_2$ between  the telescopes.  As is well known in quantum optics \citep[see e.g.,][]{Mandel}, assuming a chaotic light source \citep[which is valid for thermal astrophysical sources, see][]{Dravins2012, Foellmi2009}, one can relate the cross correlation of the intensities $\langle I_1 \cdot I_2 \rangle$ to the absolute square of the spatial correlation function $| \gamma_{12} |^2$ between the two telescopes:

\begin{equation}
    \langle I_1 \cdot I_2 \rangle = \langle I_1 \rangle \langle I_2 \rangle (1 + |\gamma_{12}|^2)
\end{equation}
Since the source is assumed to be chaotic, the intensity fluctuations will average out over timescales which are much longer than the coherence time of light. Thus, per definition: $ \langle \Delta I \rangle =  \langle I - \langle I \rangle \rangle= 0$. The cross correlation of the intensity fluctuations $\Delta I_1$ and $\Delta I_2$ will be:

\begin{equation}
    \langle \Delta I_1 \cdot \Delta I_2 \rangle  = \langle I_1 \rangle \langle  I_2 \rangle |\gamma_{12}|^2
\end{equation}

If one has a continuous source, then $|\gamma_{12}|^2$ corresponds to the correlation of photons coming from different small elements of the source's image on the sky. It will be identical to the the Fourier magnitude of the source distribution $\Sigma$:

\begin{equation}
\label{eq:gamma12}
    |\gamma_{12}|^2 = \big(\,\mathcal{F}[\Sigma]\,\big)^2.
\end{equation}

We can write down the spatial separation of the two telescopes with respect to the line of sight to the source for an optical wavelength $\lambda$, as defined in \cite{Dravins2012}: $\boldsymbol{B} = \lambda \cdot (u,v,w)$. The $w$ component accounts for the time delay due to the obliqueness of the telescope plane. Thus, the $(u,v, w=0)$ plane represents the scenario where the line of sight lies perpendicular to the plane on which the telescopes lie. Figure \ref{fig:uv_plane} demonstrates how the spatial correlation function of an elliptical source distribution of a few microarcseconds is represented in the $(u,v, 0)$ plane. For a source of angular diameter $\delta$, the characteristic scale up to which $|\gamma_{12}|^2$ is significantly non-zero in the $(u,v)$ plane is $\lambda / \delta$. In principle, one could choose to use a wavelength that best conforms to a given range of baselines and source angular size. In practice, astrophysical sources shine the brightest in a particular colour and detectors are often optimised in a single band for intensity interferometry. These effects restrict the free choice of the wavelength. Considering a $350\,$nm optical wavelength, a $5\,$km baseline will reach a resolution of about $10\,\mu$as.

The relationship between the $(u,v,w)$ vector and the baseline projections is
given by the product rotation: 

\begin{equation}\label{eq:uvw-rotation}
\begin{pmatrix} u \\ v \\w \end{pmatrix} =
\frac{1}{\lambda} \; R_x(\delta) \, R_y(h) \, R_x(-l)
\begin{pmatrix}
B_{\rm E} \\
B_{\rm N} \\
B_{\rm up}
\end{pmatrix}    
\end{equation}
where
\begin{equation}
R_x(\delta) = 
\begin{pmatrix}
1 & 0 & 0 \\
0 & \cos\delta  & -\sin\delta \\
0 & \sin\delta  &  \cos\delta
\end{pmatrix}    
\end{equation}
and similarly for $R_x(-l)$ while
\begin{equation}
R_y(h) = 
\begin{pmatrix}
 \cos h & 0 & \sin h \\
0       & 1  & 0 \\
-\sin h & 0  & \cos h
\end{pmatrix}    
\end{equation}
where $l$ is the latitude of the setup, $\delta$ is the declination and $h$ is the hour angle of the source.
Expanding out the product, Eq. \eqref{eq:uvw-rotation} is equivalent to Eq.~(7) from \citet{Dravins2012}.

\subsubsection{A hypothetical Telescope Array}
\begin{figure*}
    \centering
    \includegraphics[width = 0.95\textwidth]{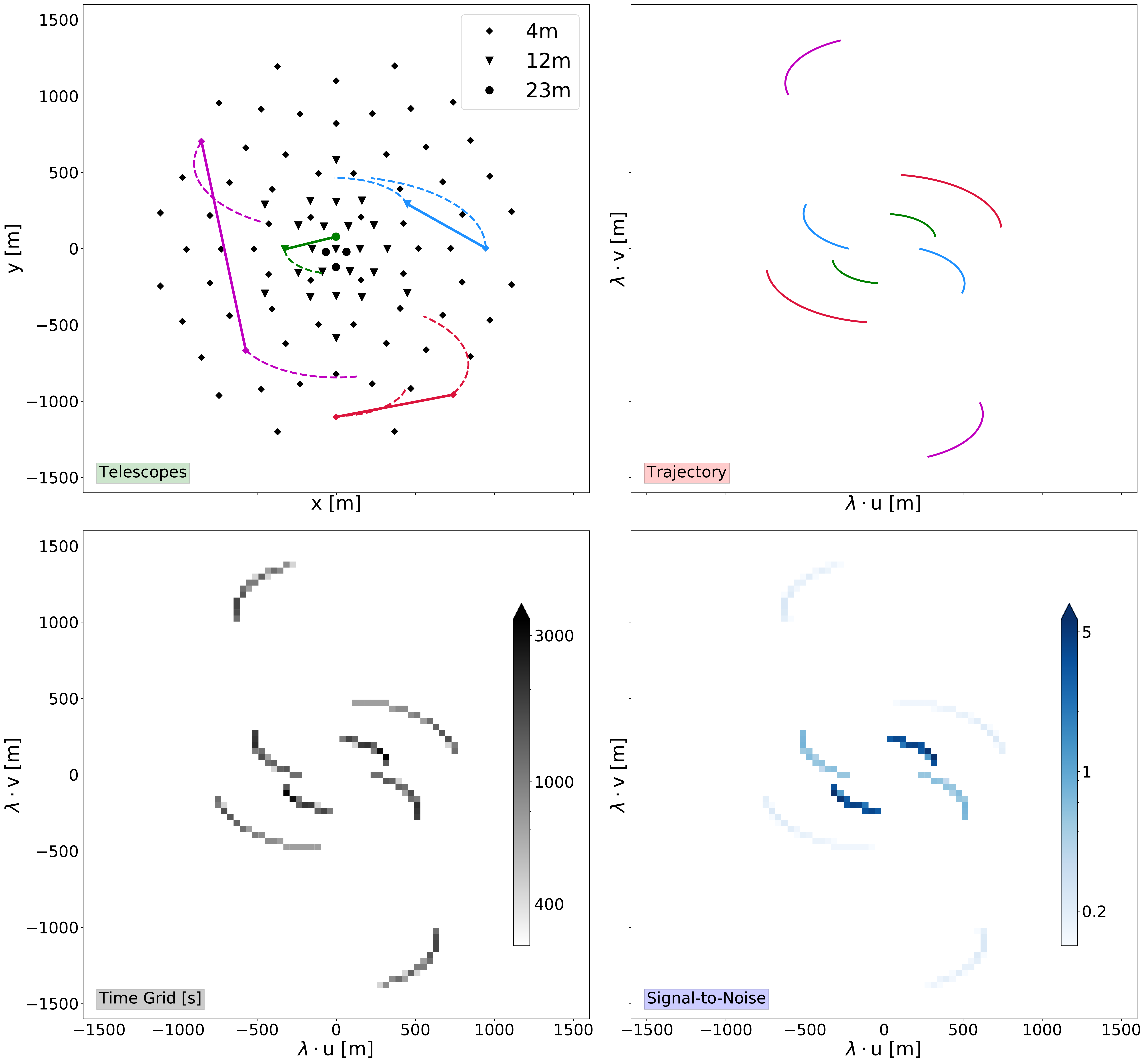}
    \caption{\textbf{Top left:} Layout of the hypothetical telescope array ACHSO in the southern hemisphere 
    Four different baselines are highlighted for demonstration purposes with solid lines with the integration time set to $\sim$5 hours. Depending on where the source is located in the sky, the Earth's rotation induces a sweeping motion on the plane. The resulting trajectories for a source with a declination of $30^\circ$ are shown with dashed lines. \textbf{Top right:} The trajectories of the selected baselines in the $(u,v)$ plane. Note that the lines here are not the same ones as the dashed lines in the top left panel, but rather they represent the sweeping of the ``baseline'' due to Earth's rotation.  \textbf{Bottom left}: The time grid of the selected baseline trajectories projected onto pixels. \textbf{Bottom right:} The Signal-to-Noise ratio on the individual pixels with the same integration time and using 1024 channels per detector (SNR $\propto \sqrt{N_{\mathrm {channel}}}$). }
    \label{fig:angle_plot}
\end{figure*}

Since intensity interferometry has already been proposed for the Cherenkov Telescope Array (CTA), we will use its telescope configuration as an example to demonstrate our method. We consider a hypothetical array of Cherenkov telescopes, the Array for CHerenkov Shower Observations (ACHSO), located at the proposed Chilean site for CTA, with the same telescope layout and subsequent number of baselines available. This comprises 4 large-sized telescopes (23\,m), 25 medium-sized telescopes (12\,m) and 70 small-sized telescopes (4\,m) \citep[see][]{CTAconcept2013}. The array layout can be seen in top left of Figure \ref{fig:angle_plot}.  The total number of baselines in this case turns out to be: $n(n-1)/2 = 4851$, where $n=99$ is the total number of telescopes in the array. However, for the purposes of this measurement we assume a considerably improved optical performance (point-spread-function, PSF and f-number) over that typical for IACTs. This renders ACHSO a much more suitable array for optical intensity interferometry.
In the results sections we will also explore various combinations of the ACHSO,  Very Large Telescope (VLT) and Very Large Telescope (ELT). Within our assumptions, applying our method to different arrays simply requires adjusting the baselines and the collecting areas of the telescopes.  

Arrays of Cherenkov telescopes are gamma-ray facilities designed to detect the Cherenkov emission produced by Extensive Air Showers generated by gamma rays entering the Earth's atmosphere. Imaging Atmospheric Cherenkov Telescopes (IACTs), are designed to detect these faint Cherenkov flashes - as the spectrum of Cherenkov light lies in the optical wavelength band (peaking towards the blue end), IACTs are sensitive optical instruments, with large mirror reflectors. The coherent wavefront of Cherenkov radiation from Extensive Air Showers typically reaches the ground within a few ns; accordingly the standard cameras employ fast imaging techniques to capture images of the air shower. 
For intensity interferometry, however, continuous signal integration is necessary, requiring a different data acquisition process to the short readout windows used in Cherenkov observations. Currently, options to incorporate intensity interferometric capabilities are being explored, either as an alternative operation mode of the cameras or as separate equipment mounted on the lid of the cameras when they are closed \citep{Dravins2012}.
Existing IACT arrays have recently made astrophysical measurements, demonstrating the feasibility of conducting intensity interferometry with a removable plate that can be easily added to the Cherenkov camera focal plane \citep{matthews,Kieda}.
Intensity interferometry with IACTs works well given their comparatively good mirror reflectivity, of typically $> 85\%$ in the optical wavelength range \cite{Gaugmuon}. However, the sensitivity towards $\sim 12$\,mag objects is only achievable with the improved optical properties provided by ACHSO.
With regards to the PSF, we estimate that in order to reduce the background noise contribution to a level significantly below that of a mag $\sim12$ source, the PSF of ACHSO needs to be improved by a factor $\sim 100$ over current IACT instrument performance. This is because, when observing under moonlight, the background brightness amounts to $\sim 17$ magnitudes per square arcsecond. Alternatively, the source could be observed under dark sky conditions, when the background contribution per square arcsecond is $\sim 5$ magnitudes less. This requirement can, however, be alleviated somewhat by the use of a narrowband filter, to reduce the broadband NSB contribution to the signal in the wavelength range of interest. 

As the Cherenkov light from Extensive Air Showers is particularly faint, observations are sensitive to background light from stars and can be severely affected by moonlight, to the extent that observations typically do not take place when the moon is illuminated at high percentages or high above the horizon. Although disadvantageous for gamma-ray observations; it may be feasible to employ this time for intensity interferometry, unless otherwise needed for calibration purposes \citep{Dravins2012}.
Nevertheless, the overall technological adjustments are comparatively minor and non-invasive to the primary purpose of IACT arrays as gamma-ray facilities. \cite{Dravins2012} outline and discuss the adjustments necessary to successfully operate as an optical interferometric instrument. 

\subsubsection{Signal-to-Noise Estimation}
\label{subsec:StN}

The Signal-to-Noise ratio in interferometry has an interesting
dependence on the source and instruments.  Let $n$ be the spectral
photon flux of the source (photons $\rm m^{-2}\,s^{-1}\,Hz^{-1}$) and
let $|\gamma_{12}(\mathbf{r})|^{2}$ be the normalised correlation.
Let $A$ be the area of each light bucket (the geometric mean if not
equal), $\Delta \nu $ the observing optical bandwidth, $\alpha$ is the
quantum efficiency of the photon detectors, $\Delta f$ the electronic
bandwidth (i.e., reciprocal of the resolution time), and $\Nt=\Delta f\times T$ (where $T$ is the integration time) is the number of time slices used to measure the signal.  Then:
\begin{equation}
\label{eq:snr formula}
\bigg(\frac{S}{N}\bigg)_{RMS}=A  \:\alpha \: n \:|\gamma_{12}(\mathbf{r})|^{2} \: \sqrt{\frac{\Nt}{2}}.
\end{equation}
for a single spectral channel \citep[see e.g.,][]{bohec}.  A more
detailed expression is given by \cite{hbt1957} in their Eq.~(3.62).
We can express the signal amplitude as:
\begin{equation}
S  = (A \: \alpha \: n)^{2} \:|\gamma_{12}(\mathbf{r})|^{2} \: \frac{\Delta\nu}{\Delta f} \: \Nt
\label{eq:signal amplitude}
\end{equation}
whose interpretation is the expected excess of correlated photons in time $T$.  The noise amplitude is:
\begin{equation}
N = A \: \alpha \: n \: \frac{\Delta \nu}{\Delta f} \: 
\sqrt{2\Nt},
\label{eq:noise amplitude}
\end{equation}
and comes from the shot noise in the number of correlated photons.
Both $S$ and $N$ here are dimensionless, and differ from the corresponding expressions in \cite{hbt1957} by a dimensional factor.

The quantum efficiency of the photo-multiplier-tubes (PMT) at the CTA are around 40\% at 350\thinspace nm wavelength, whilst the SiPMs used by the smallest telescopes can have an efficiency of up to 50\% at 470 nm \citep{sakurai,heller}. Recently, \citet{wollman} report a 1024-channel detector using superconducting nanowires with a time resolution better than $0.5\,$ns ($\Delta f \sim 2 \times 10^9 $Hz). The detection efficiency of these is only up to 23\% so far, but is expected to improve considerably as the technology is devoleped further.  As part of ACHSO, we consider that such nanowire detectors can in principle be installed on all telescopes. Moreover, new technologies may enable more channels with better time resolution and more efficient photon detectors might be developed prior to the launch of LISA. We further assume that the main mirrors are isochronous to better than the detector time-resolution.

\subsubsection{Parameter fitting}
\label{subsec:signalrec}
We will now demonstrate that it would be possible to deduce the orientation of the source in the sky from the variations of $|\gamma_{12}(\bold{r})|^2$ over the $(u,v)$ plane using intensity interferometry. \cite{hanbury} already applied the basic idea applied to the rotational flattening of Altair (see his \S11.7).  In the present work we have developed a fitting technique that combines the data collected by all the available ACHSO pairs of telescopes. Due to the rotation of the Earth, each baseline moves in the  $(u,v)$ plane. Knowing the position of the binary in the sky and the coordinates of ACHSO, we can compute the trajectories of the baselines using Eq. (\ref{eq:uvw-rotation}). They cover the $(u,v)$ plane in an inhomogeneous way as shown in the top left corner of Figure \ref{fig:angle_plot}. We discretise the problem and divide the $(u,v)$ plane into small pixels with area $\rm\approx 40\,m\times40\,m$. Considering one night of observation time, we account for the fraction of time each pair of detectors spends collecting data in each pixel. We then obtain an effective integration time $t_{ij}$ at each grid point in the $(u,v)$ plane. Thus

\begin{equation}
\sum_{i,j} t_{ij} =  N_B \cdot T,
\end{equation}
where $N_B$ is the number of pairs (baselines) and T the observation time. This procedure holds only if the area of the pixels we choose is at least as big as the smallest telescope of the detector. The Signal-to-Noise Ratio (SNR) adds in quadrature, therefore we end up with an effective SNR for each pixel that takes the simple form of Eq. (\ref{eq:snr formula}) using $t_{ij}$ as a local integration time and the mean area of the telescopes that eventually crossed the pixel as the effective collecting area.
We then perform a likelihood analysis of a mock signal with known orientation $\phi_0$ to estimate our capacity to reproduce the true value.

We first generate a signal for a given layout of telescopes, using Eq. (\ref{eq:signal amplitude}). In order to model a realistic measurement, this noiseless signal must be completed with noise characterised by Eq. (\ref{eq:noise amplitude}). Given the source flux of  CD$-30^\circ$ 11223, and the layout of the ACHSO; a baseline consisting of telescopes with $4$m diameter, located in the $(u,v)$ plane such that $\gamma=0.5$, will receive approximately $16$ coincidence photons per hour. 

The number of coincident photons pairs that arrive on two detectors is a Poisson process. However, since the number of events detected during a significant integration time with each baseline will be large enough, it is acceptable to model the noise with a Gaussian distribution. This $S+N$ constitutes the mock data $s_{ij}^\phi$, from which we aim to fit the corresponding angle $\phi$. We then generate a set of noiseless signals that span the full range of possible orientations. For a test signal with orientation $\theta$, assuming a flat prior, the total posterior distribution is the product of the likelihood of each pixel: 

\begin{equation}
\label{eq:posterior}
P(\theta \vert \phi ) = \prod_{i,j} \mathcal{L}_{ij}(\phi  \vert \theta ) = \prod_{i,j} \frac{1}{2\pi \sigma^2_{ij}} \exp\left[ -\frac{\left( s_{ij}^\theta - s_{ij}^\phi \right)^2}{2\sigma^2_{ij}}\right]
\end{equation}
where $\sigma_{ij} \propto t_{ij}^{-1/2}$ is the standard deviation of $s_{ij}^\phi$, obtained from Eq.~\eqref{eq:noise amplitude} using the effective integration time.
The most likely value of $\theta$ will tend to the true value as a function of observation time.

\section{Binary System CD$-30^\circ$ 11223}\label{sec:binary}
As a proof of concept, we want to apply the methods presented in Section \ref{sec:methods} specifically to the brightest LISA confirmation binary system, CD$-30^\circ$ 11223.
\begin{figure*}
    \centering
    \begin{subfigure}[t]{0.49\textwidth}
        \hspace*{0.1cm}
        \centering
        \includegraphics[height=9.cm]{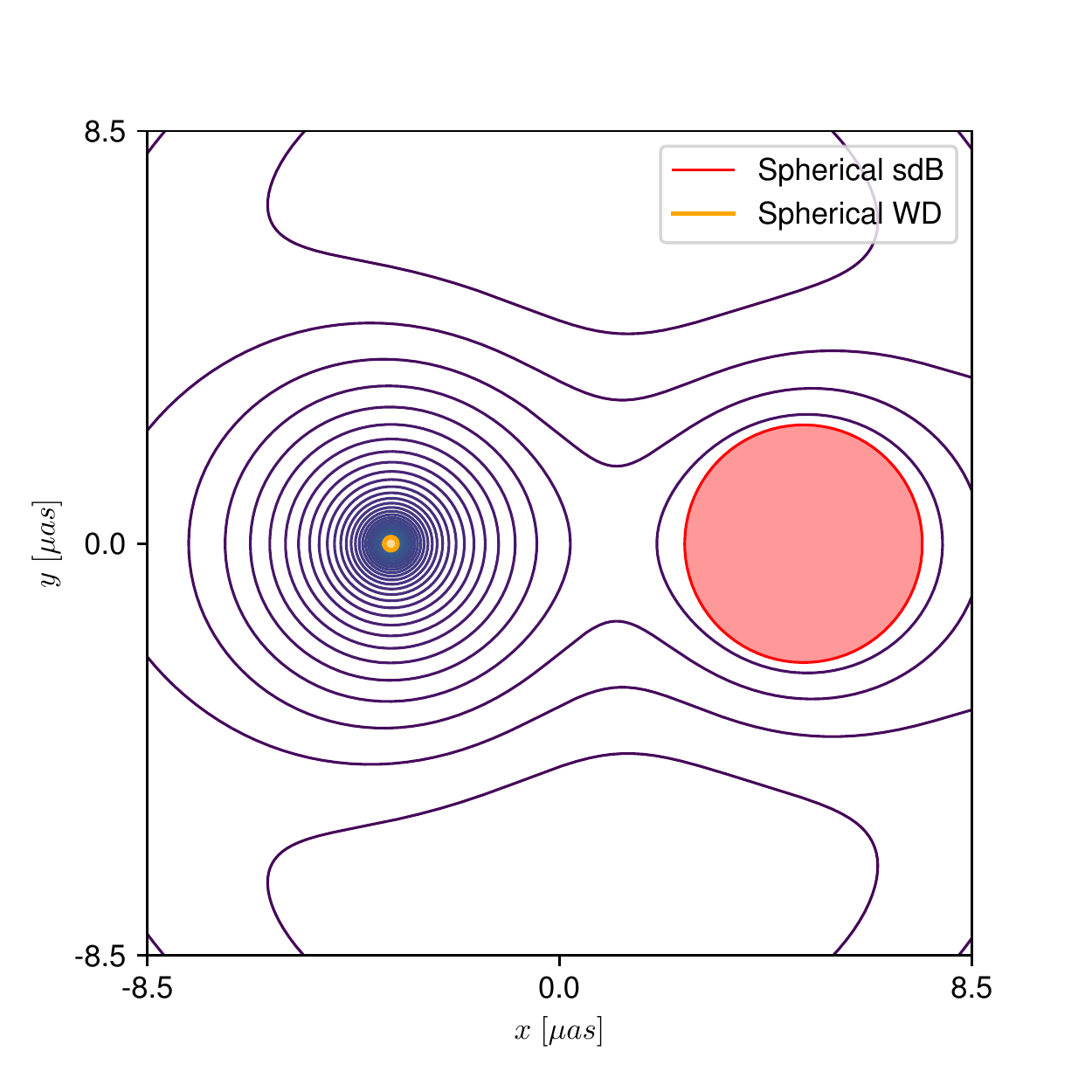}%
    \end{subfigure}%
    ~ 
    \begin{subfigure}[t]{0.49\textwidth}
        \vspace*{-9.02cm}
        \hspace*{-0.2cm}
        \centering
        \includegraphics[height=9.1cm]{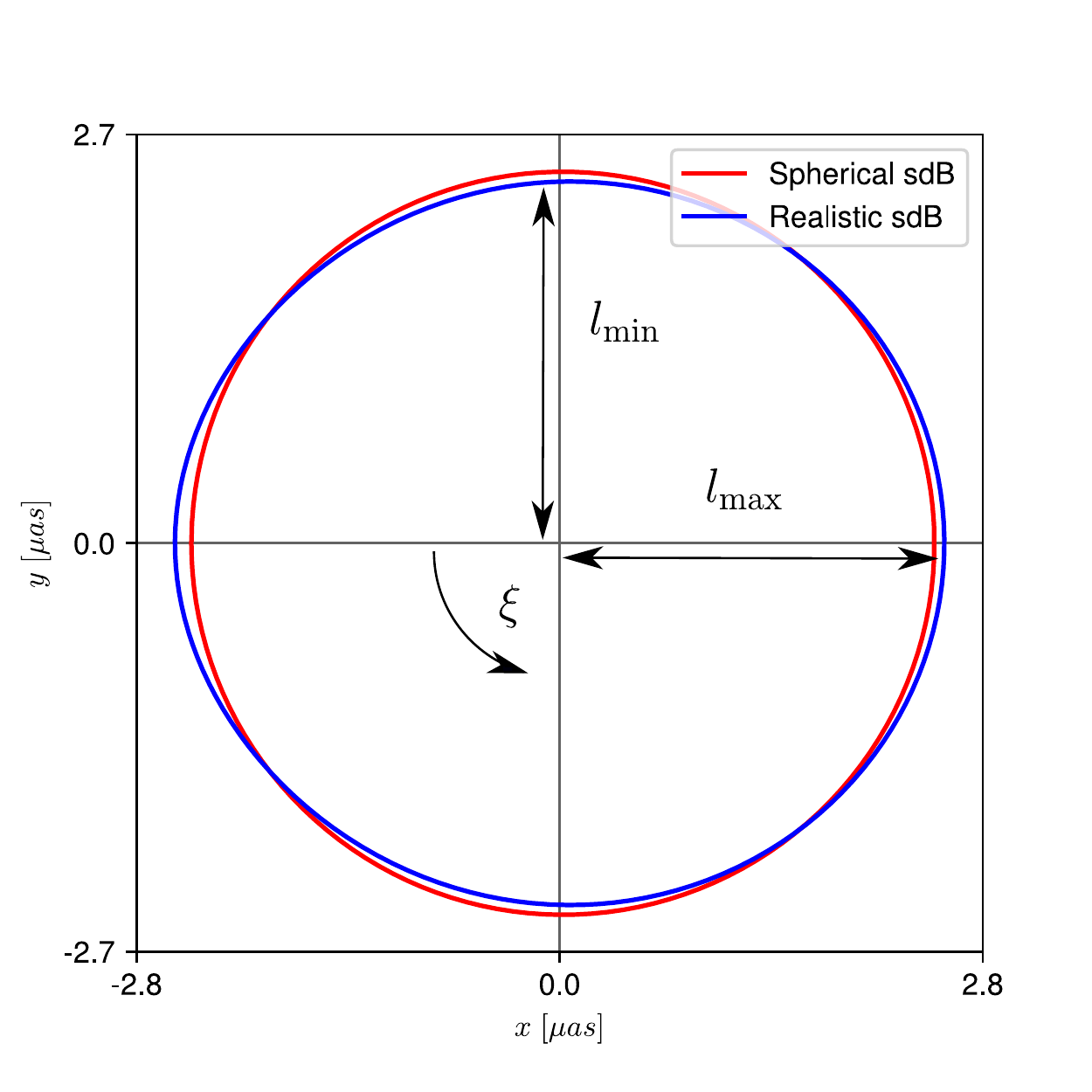}%
    \end{subfigure}
    \\
    \vspace*{-0.41cm}
    \begin{subfigure}[t]{1.0\textwidth}
        \hspace*{-0.3cm}
        \centering
        \includegraphics[height=6.2cm]{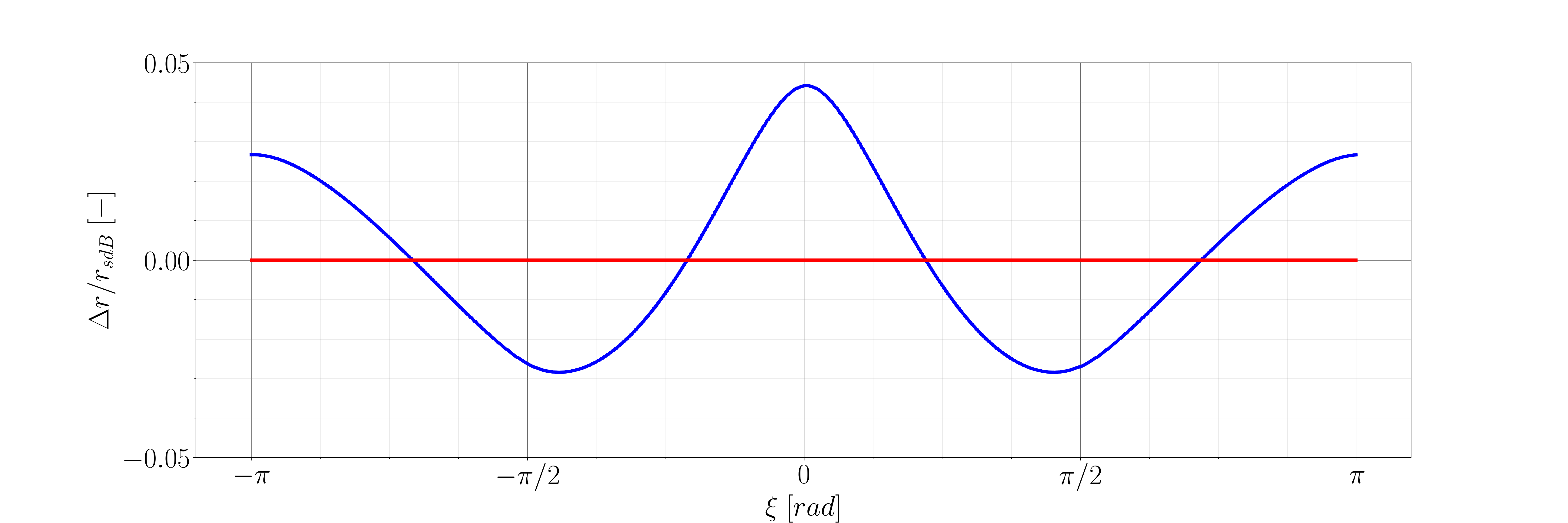}%
    \end{subfigure}
    \caption{\textbf{Top left:} Representation of the Roche equipotential lines for the binary system CD--30$^\circ$ 11223. The spherical approximated shapes of the white dwarf (in yellow) and the sub-dwarf (in red) are shown. \textbf{Top right:} Representation of the realistic shape of the sdB deformed by the tidal forces. The major and minor semi-axis, $r_{\rm min}$ and $r_{\rm max}$, are presented for the spheroid approximation. $\xi=0$ indicates the direction towards the WD. \textbf{Bottom:} Relative deformation from the spherical approximation computed as $\Delta r/r_{\rm sdB}$ as a function of the angle $\xi$.}
    \label{fig:RocheApproach}
\end{figure*}

\begin{table}
\centering
 \begin{tabular}{||l l l l||}
 \hline
 sdB temperature  & T$_\text{eff}$ & $29.2(4)\cdot 10^{3}$ & $[K]$\\
 Orbital period  & P & $1.175497738(40)$ & $[\text{hours}]$\\
 Parallax $^\dag$ & $d$ & $2.963(80)$ & $[\text{mas}]$\\
 \hline
 \bf{Solution 1} &&&\\
 sdB mass  & m$_\text{sdB}$  & $0.47(3)$ & $[M_\odot]$\\
 sdB radius & R$_\text{sdB}$ & $0.169(5)$ & $[R_\odot]$\\
 WD mass  & m$_\text{WD}$ & $0.74(2)$ & $[M_\odot]$\\
 WD radius & R$_\text{WD}$ & $0.0100(4)$ & $[R_\odot]$\\
 Separation  & $a$ & $0.599(9)$& $[R_\odot]$\\
 Orbital inclination  & $\mathcal{I}$ & $83.8(6)$& $[^\circ]$\\
 \hline
 \bf{Solution 2} &&&\\
 sdB mass  & m$_\text{sdB}$  & $0.54(2)$ & $[M_\odot]$\\
 sdB radius  & R$_\text{sdB}$ & $0.179(3)$ & $[R_\odot]$\\
 WD mass & m$_\text{WD}$ & $0.79(1)$ & $[M_\odot]$\\
 WD radius  & R$_\text{WD}$ & $0.0106(2)$ & $[R_\odot]$\\
 Separation  & $a$ & $0.619(5)$& $[R_\odot]$\\
 Orbital inclination & $\mathcal{I}$ & $82.9(4)$& $[^\circ]$\\
 \hline
 \end{tabular}
\caption{Parameters of the binary system CD$-30^\circ$ 11223. The two solutions are obtained with ISIS and Goodman, a spectrograph at the SOAR telescope, and allow to investigate systematic errors. However only solution 2 is used in this study \citep{Geier2013}, $^\dag$\citep{kupfer}.}
\label{table:parameters}
\end{table}

This system hosts a white dwarf (WD) and a hot helium subdwarf (sdB) orbiting each other in a binary fashion. As summarised in Table \ref{table:parameters}, the configuration of the system is such that the two bodies are close with a comparatively short orbital period, making it suitable for LISA's verification tests \citep{kupfer}. Moreover, the total effect of gravitational interactions paired with large orbital velocities generates extreme tidal forces which stretch the subdwarf. Therefore the configuration of CD$-30^\circ$ 11223 makes it a promising candidate to infer the angle $\phi$ of the system from  intensity interferometry within a reasonable amount of observational time. 

In the following we apply the approaches described in Section \ref{subsec:tidal_stretching} to estimate the induced deformation of the sub-dwarf. 
The Roche potential offers an analytic description for the gravitational potential of two tidally locked, co-rotating bodies on circular orbits with common orbital period $\omega$. 
Since the orbital period of CD$-30^\circ$ 11223 is significantly smaller than 10 days, the assumption of a circular orbit and tidal locking for the sdB is justified according to \citet{Zahn1989}.
We thus implement the pipeline using the classical Roche potential presented in Section \ref{subsec:tidal_stretching} as both conditions are fulfilled for the CD$-30^\circ$ 11223 system.
However, we note that one could adopt a more generalised Roche potential as presented by \citet{Avni1982} to relax some of the restrictive orbital assumptions in order to model binary systems with more extreme cases of misalignments.\\

On the top left of Figure \ref{fig:RocheApproach} we show the equipotential lines computed numerically given the parameters of the second solution of the binary listed in Table \ref{table:parameters}, while on the top right we zoom on the subdwarf and show the actual deformation caused by tidal forces. On the bottom we show the radial deformation, given by the ratio between the radius at azimuthal angle $\xi$ and the mean radius $r_{\rm{sdB}}$. We see that the deformation function does not precisely follow a cosine, since the true shape of the sdB will be a Roche lobe. However, to first order, the spheroidal approximation holds.  
We can therefore compute the ratio $\kappa$ between the minor and major semi-axes, $\kappa = r_{\rm min}/r_{\rm max}$. From the two solutions of the binary we obtain the following numerical results:
\begin{equation}
    \kappa_1 = 0.940(1)  \\
    \kappa_2 = 0.940(3) \label{eq:kestimate1}
\end{equation}
As a confirmation, we apply the brightness variation method described in Section \ref{subsec:tidal_stretching} to the V-band light curve data presented in \citet{Geier2013}, Figure 5, where the major fluctuations are precisely due to tidal effects. With the data available we estimate:
\begin{equation}
\kappa = \frac{I_{e, \Omega}^{\mathrm{min}}}{I_{e, \Omega}^{\mathrm{max}}}  = 0.923(3) \label{eq:kestimate2}
\end{equation}
which is in the same range as the result obtained with the more sophisticated numerical approach presented above. We would like to make the reader cautious about the fact that the errors presented in Eqs.\,(\ref{eq:kestimate1}),\,(\ref{eq:kestimate2}) are only due to measurement uncertainties. Therefore, the difference between the values of the two estimates comes from the different assumptions made for the two methods, which we discuss in what follows.

The calculated values of $\kappa$ are subject to different error sources. The Roche potential approach suffers mainly from uncertainties related to the observed properties of the binary system. Although smaller contributions are expected to arise from potential perturbations caused by the star deformations. In fact we compute the potential assuming point mass objects. In this particular case these effects are negligible, since the shape of the white dwarf is essentially spherical and not perturbed by the sdB. In general this method is reliable for sufficiently precise measurements of its principle properties as the mass, radius and separation of the stars. 

Concerning the variation in radiant intensity in the direction of the Earth, the uncertainty in the estimation of $\kappa$ is of the same order of magnitude as the Roche potential method, given the available data. However, this method is based on a simplistic physical model and therefore ought be considered approximate. Firstly, it relies again on the accuracy of the measured flux. Secondly, relativistic Doppler boosting effects can vary the flux maxima and minima during the system's revolution. However, as we can seen from Fig.\,5 of \citet{Geier2013}, such an effect is negligible with respect to the variations due to the spheroidal deformation.  Thirdly, the body shape is assumed to be spheroidal and its radiative spectrum follows the one of a perfect black body. Furthermore the effect of limb-darkening on the  flux, and therefore on the estimated area, is neglected.
In conclusion, this method is rather approximate and should be used solely as an order-of-magnitude estimation.

\section{Results}
\label{sec:results}
\subsection{Observation Time Calculations}

We consider ACHSO in a scenario corresponding to the planned configuration for CTA (South), with 99 telescopes (4851 different
baselines) spread over $4\rm\,km^2$.  We assume all the telescopes are
able to observe in 1024 spectral channels with a time-resolution of
$0.5\rm\,ns$ \citep[cf.][]{wollman} and that the background and all
systematic effects are much less than the counting noise
(Eq.~\ref{eq:noise amplitude}).  For simplicity we also assume that
all hour angles are accessible over the course of a year (in effect,
taking the source to be circumpolar); this is not true for
CD$-30^\circ$~11223 at the site latitude, but is not expected to
significantly change our estimates.

We generate a noisy $|\gamma_{12}|$ with Eq.~\eqref{eq:gamma12} given
the stretched geometry of the star, assuming the true binary
orientation corresponds to $\phi=0$.  We then compute the posterior
probability (Eq.~\ref{eq:posterior}) for inferring $\phi$.
Figure~\ref{fig:all_likelihoods} shows the posterior for the binary
orientation for various observation times, ranging from one week to
twelve months,\footnote{For a month, we assume $30\times12\rm\,hr$ of observing time.} using the ACHSO. A measurement of the binary orientation with an uncertainty of $\pm$5$\degree$ at $1\sigma$ confidence level, would require the full ACHSO layout for about one month. Admittedly, this does not seem very promising, but could still be an interesting constraint for the polarisation predictions (see Figure \ref{fig:GW_error_plot_amplitudes}). In Figure \ref{fig:GW_error_plot_obstime} we show how the constraints on the angle $\phi$ get better with more observation time.

Although we can assume that the background light is insignificant for ACHSO, this is unfortunately
unlikely to hold for the true CTA because the expected point
spread function (PSF) would let in significant amounts of background light from
the sky.  The PSF has two origins.  One is spherical aberration and so
on from the mirror figure, which as an isolated effect, could in principle be corrected using
customised secondary mirrors, as was done in the Hubble Space
Telescope \citep[see Figs.~1--3 in][]{1994ApJ...435L...7J}.   However, the other
contribution to the PSF is the roughness of the mirror surfaces,  which
cannot be compensated for  and is a limiting factor that also prohibits the use of secondary optics to correct for spherical aberrations.  Mirrors for normal telescopes are polished
to sub-wavelength smoothness, but because of the cost of
optical-quality polishing, Cherenkov air telescopes polish only enough
to keep the PSF acceptable for imaging Cherenkov showers: $\approx
10^{-3}\rm\,radians$ or $\approx 3'$, that is, comparable with
naked-eye seeing \cite[see e.g.,][]{2015SPIE.9603E..07T,2013MNRAS.430.3187R}.  If the
night-sky within the PSF is comparable to or brighter than a given
source, the photon noise scales with the PSF, and as a result the
observation time scales with the area in the PSF.  For the
PSF of typical IACTs, simulations by \cite{2013MNRAS.430.3187R} indicate
that for sources beyond magnitude~10, the observing times needed become
impractical and source confusion between stars within the field-of-view also starts to become an issue. 
The PSF quality is therefore a fundamental difference between CTA and ACHSO.

\subsection{Multi-array combinations}\label{sec:vlt}
In this section we will explore two combinations of ACHSO, VLT and ELT. Considering multi-array combinations allows us to cover a significantly broader region of the $(u,v)$ plane, due to the longer base-lines. Moreover, optical telescopes such as VLT and ELT have superior mirrors and a low point spread function (PSF). We start with the combination of ACHSO and VLT.

\begin{figure*}
    \centering

    \includegraphics[width = \textwidth]{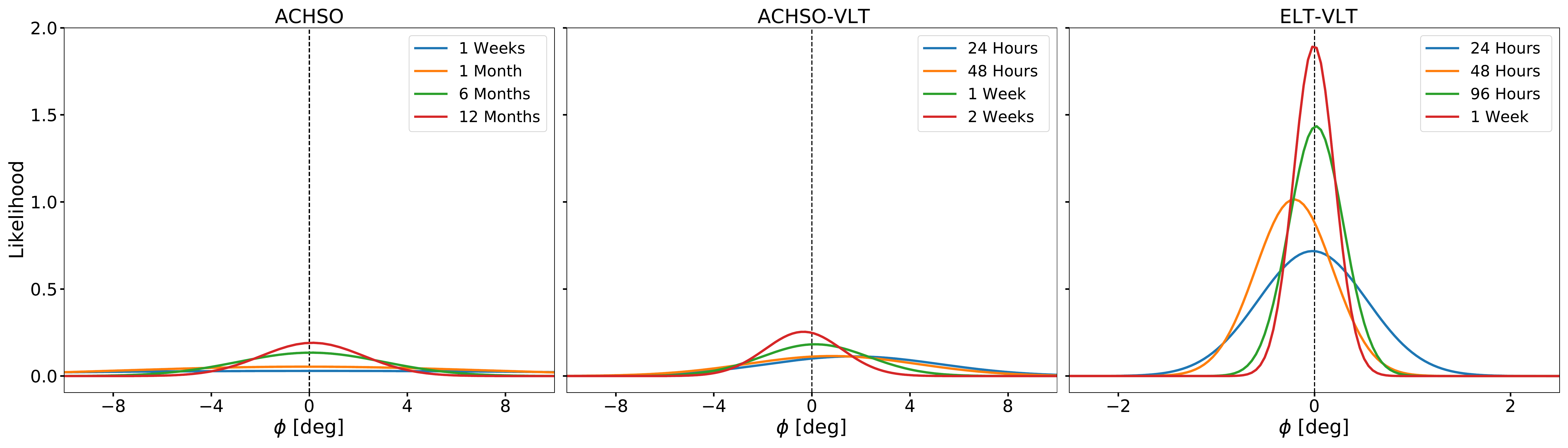}

    \caption{We plot the posterior probability distributions (PDs) of the orientation angle $\phi$ obtained by fitting different mock signals with the same assumed prior value of $\phi_0 =0^{\circ}$ (represented by the dashed line). The gaussian PDs are constructed assuming various number of nights of observations for different combinations of telescope arrays: ACHSO only (\textbf{left panel}), ACHSO-VLT (\textbf{central panel}), ELT-VLT (\textbf{right panel}).}
    \label{fig:all_likelihoods}
\end{figure*}

The VLT is made of four 8\thinspace m telescopes, in the Atacama desert, $\sim11\rm\,km$ away from ACHSO. We assume the same quantum efficiency and bandwidths for both arrays and carry out the analysis proposed in section \ref{subsec:signalrec} including all the baselines between the pairs of telescopes. Figure \ref{fig:all_likelihoods} shows the likelihood estimation of the binary orientation for various observation times using ACHSO combined with the VLT. We obtain drastically better results for the estimation of $\phi$. As an example, the uncertainty in the measurement falls down to $\pm3^\circ$ at 2$\sigma$ in just one week of observation. In Figure \ref{fig:GW_error_plot_obstime} we show how the constraints on the angle $\phi$ get better with more observation time.

Even more powerful is the combination of VLT and the Extremely Large Telescope (ELT). The ELT is made of a 39.3\thinspace m telescope, in the Atacama desert, $\sim21\rm\,km$ away from ACHSO. We assume the same quantum efficiency and bandwidths for it, and carry out the analysis proposed in section \ref{subsec:signalrec} for the four baselines between the ELT and VLT telescopes.
 Figure \ref{fig:all_likelihoods} shows the likelihood estimation of the binary orientation for various observation times. Due to the long baselines and the large collecting areas of VLT and ELT, we obtain an extremely precise constraint on the orientation angle $\phi$ in a very short observation time: an uncertainty of $\pm 1 ^{\circ}$ at 2$\sigma$ is reached in roughly 24 hours of observations, while the same uncertainty at $5 \sigma$ is reached in roughly 5 nights (60 hours) of observation.  

\subsection{Polarisation Prediction for CD--30$^\circ$\,11223}
At this point, we are able to estimate the precision of a polarisation prediction given the findings of previous sections. Here, we focus on presenting  and discussing the results for the binary system CD$-30^\circ$ 11223. Derivations for all the expressions that will be used are given in Appendix~\ref{appendix:GW}.
The ``plus'' and ``cross'' polarisations of gravitational waves emitted by CD$-30^\circ$ 11223 change the length of the photon path by $\Delta L_+(t)$ and $\Delta L_\times(t)$ respectively. The numerical values of their amplitudes depend on the position of LISA relative to the binary system (which is described through the angles $\phi$ and $\psi$ illustrated in Figure \ref{subfig:GW_LISAangles}) and are given by:

\begin{subequations}
\begin{align}
   \frac{\Delta L_+}{L} &\approx 4.2 \times 10^{-22}  \cos (2 \phi ) \\\
    \sigma_{+}&\approx 2 \sqrt{ 0.02^2 + \Delta_{\phi}^2 \tan(2 \phi)^2}\\\
    \frac{\Delta L_\times}{L} &\approx -1.0 \times 10^{-22} \sin (2 \phi ) \\\
    \sigma_{\times} &\approx 2 \sqrt{0.03^2 + \Delta_{\phi}^2\cot(2 \phi)^2}\\\
   \frac{\Delta L_+}{\Delta L_\times}&\approx - 4.1 \times \frac{\cos(2 \phi)}{\sin (2 \phi)}
\end{align}
\end{subequations}
The polarisation amplitudes depend on the unresolved orbital orientation angle $\phi$ periodically. The relative uncertainties $\sigma_{+}$ and $\sigma_{\times}$ can be multiplied by the amplitudes to give the absolute error of the polarisation prediction. They depend partly on observational errors listed in Table \ref{table:parameters} and partly on the error in the orientation angle $\phi$. In Figure \ref{fig:GW_error_plot_amplitudes}, we have shown the full range of possible polarisation amplitudes and uncertainties given by different errors in the orientation angle. A convenient way to express the total accuracy of a polarisation prediction is the square sum of the relative uncertainties:
\begin{equation}
    \mathcal{R}^2 = \sigma_+^2 + \sigma_\times^2 \approx 0.08^2 + \frac{40}{3} \Delta_{\phi}^2
    \label{eqn:GW_mean_error_max}
\end{equation}
Here $\Delta_{\phi}$ is the error in the measurement of the angle $\phi$, while the fixed numerical value of $0.08^2$ is the combined uncertainty arising from the measurements given in Table \ref{table:parameters}. Briefly, $\mathcal{R}$ is a measure of the total error that can be expected from all the uncertainties in the observational data about the binary system (Table \ref{table:parameters}), along with the error in the angle $\phi$ arising from the intensity interferometry approach. As seen in previous paragraphs, $\Delta_{\phi}$ can be reduced with longer observation times or with more ambitious telescope array combinations. The errors arising from the parameters in Table \ref{table:parameters} might also decrease as more precise measurements are published in the near future. In Figure \ref{fig:GW_error_plot_obstime}, we show the uncertainty of the polarisation prediction for the binary system CD--30$^\circ$ 11223 with the available measurements as a function of observation time for intensity interferometry. Interestingly, just a few dozen nights of observation with the VLT + ACHSO or a single day with the VLT+ELT combination essentially reduce the error in the orientation angle $\phi$ to zero.

\section{Discussion \& Conclusion} \label{sec:discussion}
\begin{figure}
    \centering
    \includegraphics[scale=0.55]{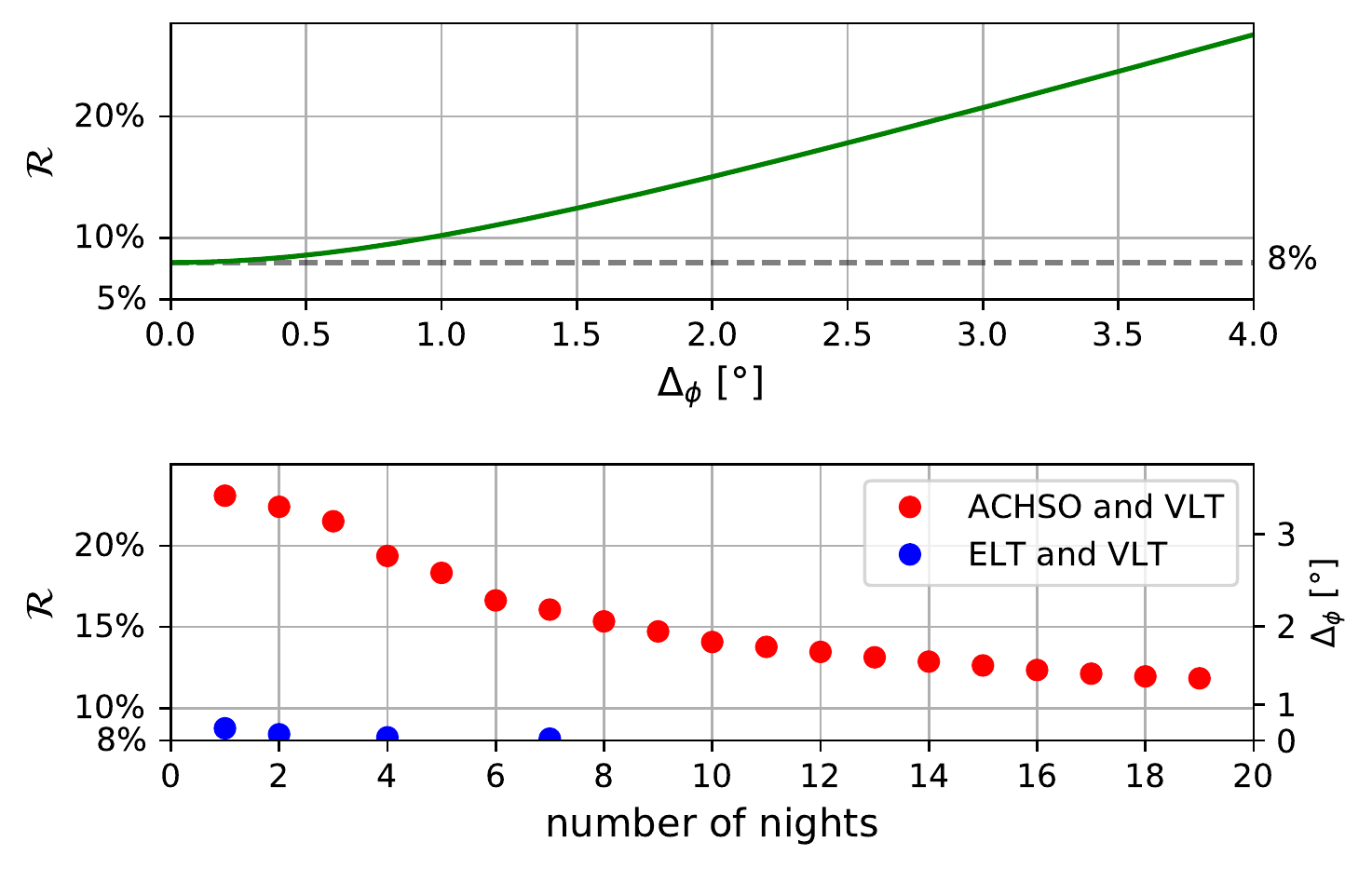}
    \caption{Here we plot the uncertainty $\mathcal{R}$ (Eq.~\ref{eqn:GW_mean_error_max}) of the polarisation prediction both as a function of $\Delta_{\phi}$ (top panel) and of observation time (bottom panel), where $\Delta_{\phi}$ is the 1$\sigma$ uncertainty in the inferred angle. From the upper panel, one sees that the mean square uncertainty increases rather sharply for an error $\Delta_{\phi}$ larger than a few degrees. The dashed 8\% uncertainty line is due to measurement errors listed in Table~\ref{table:parameters}. The lower panel shows that the combination of ACHSO and VLT can essentially remove the error due to the measurement of the orientation angle in a dozen observation nights, while the ELT+VLT take only a fraction of an observation night to reach 1 $\sigma$ confidence levels.}
    \label{fig:GW_error_plot_obstime}
\end{figure}

This paper presents a method to resolve the orientation of a LISA verification binary and thereby produce a prediction for its gravitational-wave amplitudes. The idea is to resolve the elongation of the tidally stretched star via intensity interferometry, in order to deduce the rotation axis of the binary system, which in turn determines the polarisation amplitudes of the GW. We look at a promising binary system: CD$^\circ$ 30-11223, comprised of a hot sub-dwarf and a white dwarf.\footnote{The sub-dwarf being actually gigantic compared to the white dwarf.} We deduce the elongation of the stretched sub-dwarf via two methods: using Roche potential and using radiant flux variations. Assuming an incoherent source we compute numerically the intensity interference measurable on the Earth's surface. As the Earth revolves around itself, the baselines move over the interferometric $(u,v)$ plane inhomogeneously. We compute the effective time spent collecting data (counting coincidence photons) at each point of the $(u,v)$ plane. This yields an effective signal to noise ratio, which we use to recover the orientation of the ellipticity. For a mock signal induced by an angle $\phi=0$, we compute the posterior distribution of the binary inclination angle given the mock signal and noise.
Our estimates for the binary system CD$-30^\circ$ 11223, as shown in Figures \ref{fig:GW_error_plot_obstime} and  \ref{fig:GW_error_plot_amplitudes}, suggest that the intensity interferometry approach using ACHSO can measure the orientation angle $\phi$ to a few degrees ($\sim 5^{\circ}$) in a large but not unimaginable amount of observation hours ($\sim  600\,$hr). However, using the combination of ACHSO and VLT would achieve the same accuracy in less than a tenth of the time. For the case of VLT and ELT, only a single night of observation would be required to constrain the uncertainty to roughly $1^{\circ}$ at $2 \sigma$ confidence level.

There are several considerations that might increase the prospects of this methodology in the years leading to the launch of LISA. As an example, the signal to noise ratio increases linearly with the quantum efficiency of the photon detectors and also increases as the inverse square root of their resolution time. Both of these technologies are likely to improve, thus reducing the time needed to achieve the desired precision. Measurements of the orbital parameters of verification binaries are also likely to improve significantly in coming years, further reducing the uncertainties in the polarisation prediction.
These points argue that intensity interferometry will be an essential piece of the ``multi-messenger puzzle'' needed to predict and measure gravitational waves.

There are however, a number of caveats that were not included in our estimates.

On the experimental side, we are assuming the viability of some technologies that are currently under development.  The most difficult challenge at telescopes would be the deployment of kilo-pixel photon counters.  This requires dispersing light so that each pixel gets a different narrow wavelength band.  Precise wavelength calibration is not required, just reproducibility in the sense that corresponding pixels in different detectors get the same wavelength band.  The basic idea has been discussed in the context of intensity interferometry \citep[see Fig.~3 in][]{2016SPIE.9907E..1WH}. Even though the implementation of such technology on current telescope arrays seems ambitious, it does not have to be implemented now, but rather in a dozen or so years. In our estimation, this is a reasonable proposal since prototypes already exist. In a sense, LISA itself is the most ambitious piece of technology that we are assuming in this work.

\begin{figure}
    \centering
    \includegraphics[width = \columnwidth]{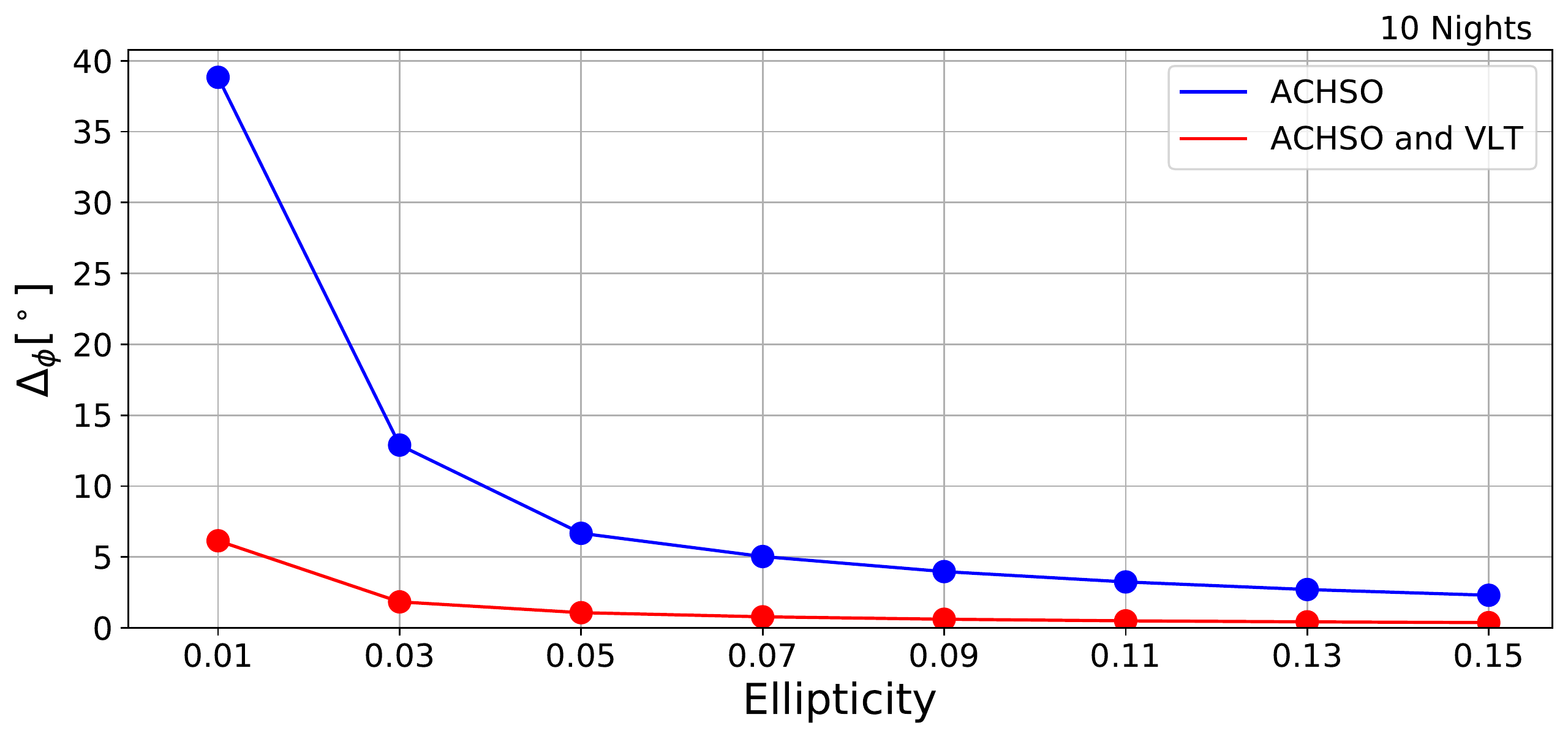}
    \caption{We plot the remaining uncertainty $\Delta_\phi$ in the orientation angle $\phi$ after ten nights of observation for a spheroidal source with the same parameters as CD$-30^\circ$ 11223, but with different ellipticities. The blue points are produced using ACHSO and the red points using the combination of ACHSO and VLT. As expected, strongly elliptical sources can be resolved more quickly.}
    \label{fig:ell}
\end{figure}

On the side of our analysis, the method of fitting noisy mock data presented in Section \ref{subsec:signalrec} assumes that the shape of the source is fixed and homogeneous. In reality, the light profile of the source will decay towards the edges because of limb darkening. Moreover, the projection of the tidally stretched star on the sky is not always teardrop shaped for an edge-on binary such as CD$-30^\circ$ 11223. Rather, it will vary from circular to teardrop depending on the phase of the orbit. Because of the technicalities of intensity interferometry, it is not straightforward to estimate how the shape and light profile of the source affects signal to noise. To get an idea for the relevance of the source's shape, we ran several realisations of a homogeneous source with different ellipticities. We show how it affects the uncertainty on the orientation angle for 10 nights of observation in Figure \ref{fig:ell}. The 2D projection of a tidally stretched star on the sky will have an apparent ellipticity that changes direction and varies periodically. On the one hand, we might expect these factors to cut the signal by a factor of 2 to 4. On the other hand, if one were to implement a light profile and the variation of the source's shape in the fitting process, they might actually lead to more efficient recover of the orientation.

Another problem in the parameter fitting is the fact that we do not explicitly model many possible sources of noise that might affect the coincidence rate of photons. An example would be atmospheric effects, which might affect the signal in a manner roughly proportional to the attenuation of blue light in the atmosphere.


Lastly, we have chosen to analyse the binary system CD$-30^\circ$ 11223 precisely because it was a promising candidate for this methodology. There are numerous intrinsic source properties, e.g.,  radiant intensity, effective temperature, amount of tidal stretching, angular size, orientation and location on the sky, that affect the efficacy of intensity interferometry. It is not straightforward to extrapolate the effects of these parameters from one single example, with the possible exception of the radiant intensity of the source which explicitly appears in the signal to noise expression. For example, a system with Roche-lobe overflow would present a more non-circular
target to resolve with intensity interferometry. There are such
systems among the LISA verification binaries, but they are even
 fainter and smaller in angular size than CD$-30^\circ$ 11223.

Are there other possibilities for detecting the orientation angle of this system? One possibility is to investigate occultations of the binary system by asteroid belt objects \citep{asteroid}. However, while asteroid occultations of stars are common events, occultation of a given star would be exceedingly serendipitous, and moreover it is not clear what the observational signature of an elliptical source by an irregular source would be \citep[cf.][]{pre-asteroid}. Another option could be detecting the polarized light of the subdwarf, reflected by the white dwarf \citep{schmid2, schmid1}. However, the white dwarf itself is not resolvable with current technology, let alone the reflected light due to the subdwarf. Thus, we believe  intensity interferometry  is the most optimistic method.

In short, there is still plenty of work to be done to direct, refine and constrain the possibilities of the intensity interferometry approach.
The combination of ACHSO (which has considerably different optical properties and a much improved PSF than CTA) and VLT is probably potent enough to resolve the orientation angle of confirmation binaries other than CD$-30^\circ$ 11223. Even though it is ambitious, the combination can be made even more powerful using the ELT. The telescope arrays could be set up for intensity interferometry during moonlit nights and would therefore not compete with other high-priority observations. By exploiting a few days per month, it is within the realms of possibility to start constraining the polarisation predictions of many LISA confirmation binaries.

\section*{Acknowledgements}

We acknowledge support from the Swiss National Science Foundation.
This paper has gone through internal review by the CTA Consortium.

\section*{Data Availability Statement}
No new data were generated or analysed in support of this research.




\bibliographystyle{mnras}
\bibliography{example} 




\appendix
\section{Computation of the Waveform AND Error estimation}\label{appendix:GW}
In this section we briefly summarise how one goes from the generation of GWs in an astrophysical system to their detection back here around Earth. For the sake of clarity, and in view of the application presented in Section \ref{sec:methods}, we adopt some simplifying assumptions.

Firstly, we assume that the system generating the GWs is an isolated binary of compact objects. This allows us to only consider the pure signal originating from the two bodies. Secondly, we assume that the angular momentum vector of the binary's orbit does not change significantly over the period of the measurement, thus neglecting the complicated effects of spin-spin and spin-orbit coupling on orbital motion. This allows us to find a unique and convenient coordinate system in which to describe the process. We also simplify the experimental side by considering only one detector ``arm''. In other words, we imagine LISA as consisting of only two satellites exchanging photons. This allows us to not have to keep track of the relative orientation of every pair of satellites with respect to the source of GWs.
Finally, we assume perfect knowledge about the binary system's physical parameters, as well as its orientation in space with respect to our idealised LISA arm. This will allow us to calculate an unique prediction for both polarisation amplitudes. Now we are ready to compute the waveform by following the scheme reported below.

We find a convenient coordinate system among the many possible choices, as detailed in Section \ref{sec:methods}. In this paper, we align the $z$-axis with the ``line of sight direction'' from the binary's centre of mass to the midpoint of our idealised LISA arm. This choice fixes the ``$x$-$y$'' or ``transverse'' plane in which GWs will produce a measurable strain. We then align the $y$-axis with the projection of the orbit's angular momentum in the transverse plane. The chosen coordinate system is illustrated in Figure \ref{subfig:GW_inclination}. This choice will greatly simplify the the explicit construction of the mass quadrupole moment of the binary system CD$-30^\circ$~11223.

Returning to the general case, this choice of coordinates fixes a unique mass quadrupole moment for the system. It is given by a $3\times3$ symmetric matrix $\mathbf{M}$ that can be used as input for the quadrupole formalism \citep{GODTHEFATHER,schaefer} to obtain the spatial part of the linear metric perturbation tensor $\mathbf{h}$:

\begin{equation}
   \mathbf{h}(t)= \frac{2 G}{c^4 d} \Ddot{\mathbf{M}}(t_d)
\end{equation}
Here $G$ is the gravitational constant, $c$ is the speed of light, $d$ is the distance between the detector and the source and $t_d = t - d/c$ is the retarded time. The ``plus'' and ``cross'' polarisations $h_+$ and $h_{\times}$ are commonly used measures of the strain produced by the metric perturbation along the transverse plane. In these coordinates, they are found by projecting the tensor $\mathbf{h}$ in the line of sight direction and removing its trace:

\begin{equation}
\begin{split}
    h_+(t) &= \frac{ G}{c^4 d}\left( \Ddot{M}_{11}(t_d)-\Ddot{M}_{22}(t_d) \right) \\\
      h_{\times}(t) &= \frac{ 2G}{c^4 d}\left( \Ddot{M}_{12}(t_d) \right)
\end{split}
\end{equation}
We now relate the physical amplitudes to those measured by an experimental setup. In our coordinates, an idealised LISA arm consists of a thin line of length $L$ centered somewhere along the $z$~axis. The two independent polarisation amplitudes are modified by two additional factors. The first is the angle $\phi$ between the $y$~axis and the projection of the detector arm on the transverse plane. The second is the angle $\psi$ that describes the intersection of the arm with the transverse plane. These two angles are illustrated in Figure \ref{subfig:GW_LISAangles}. The former affects the relative strength of the measured polarisation amplitudes while the latter simply reduces both amplitudes if the arm is tilted with respect to the transverse plane. We apply these projections as follows:

\begin{equation}
\begin{split}
    h_+(t)  &\to \cos(2 \phi) \cos (\psi) h_+(t) \\\
      h_{\times}(t) &\to \sin(2 \phi) \cos (\psi) h_{\times}(t)
\end{split}
\end{equation}
Lastly, we consider the fact that the measurement of the strain is not instantaneous. If the detector has a length $L$, a photon travelling along the arm will take a time $L/c$ to complete it's path. To represent this fact one integrates the strain along a photon's path:

\begin{equation}
\begin{split}
    h_+(t)  &\to \frac{c}{L} \int_{t - L/c}^{t} h_+(t') dt'\equiv\frac{\Delta L_+(t)}{L} \\\
      h_{\times}(t)  &\to \frac{c}{L} \int_{t - L/c}^{t} h_{\times}(t') dt'\equiv\frac{\Delta L_\times(t)}{L}
\end{split}
\end{equation}
Note that the length of the photon path is affected by the strain as $dL=\left(1+h\right)c dt$, so $L_\text{strained}=L+\Delta L$. Now we can specialise the formulae to the case of the binary system CD$-30^\circ$ 11223.
The largest contribution to the mass quadrupole moment comes from the orbital motion of the two bodies around their centre of mass. Since CD$-30^\circ$ 11223 is an isolated binary with a very short period we can assume that is on a very low eccentricity orbit \citep{Zahn1989}. In the coordinate system defined above and in Section \ref{sec:methods}, we can describe the orbit through the separation vector $\Vec{r}(t)$ between the stars:
\begin{equation}
    \Vec{r}(t) = \begin{pmatrix}
    r_x(t) \\ r_y(t) \\r_z(t)
    \end{pmatrix} = R \begin{pmatrix}
    \cos (\omega t )\\ \sin (\omega t )\cos (\mathcal{I})  \\-\sin (\omega t )\sin (\mathcal{I})
    \end{pmatrix}
\end{equation}
Here $R$ is the radial separation, $\omega$ is the orbital frequency. We can arbitrarily set the true anomaly (phase) to zero to simplify the calculations. The mass quadrupole moment $\mathbf{M}$ for two point masses is then given by \citep[see e.g,][]{PhysRev.131.435}:
\begin{equation}
    M_{ij}(t) = \mu r_i(t) r_j(t)
\end{equation}
where $\mu$ is the reduced mass of the binary. Its magnitude is given by the reduced mass multiplied by the square radial separation ($\mu R^2$). The second largest contribution to the mass quadrupole moment comes from the tidal stretching of the sdB. It turns out, that this and all other smaller contributions are negligible. Indeed, for a spheroid with principal semi-axes $(b,a,a)$ and a mass $m_{\rm sdB}$ the magnitude of the quadrupole $M_{\rm sph}$ is given by:

\begin{equation}
    M_{\rm sph}= \frac{2}{5}m_{\rm sdB}(b^2-a^2)
\end{equation}
It can be restated in terms of the parameter $\kappa = a/b$ used in section \ref{sec:binary} and the mean geometric radius $r_{\rm gm} = \sqrt[3]{a^2b}$ of the spheroid:

\begin{equation}
    M_{\rm sph}=  \frac{2}{5}m_{\rm sdB} r_{\rm gm} ^2\left( \frac{1 - \kappa^2}{\kappa^{4/3}} \right)
\end{equation}
The ratio between the sdB's quadrupole $M_{\rm sph}$ and the orbit's quadrupole $M$ is very small for our system's parameters:
\begin{equation}
    \frac{M_{\rm sph}}{M}= \frac{2 m_{\rm sdB}}{5 \mu} \frac{r_{gm^2}}{R^2} \left( \frac{1 - \kappa^2}{\kappa^{4/3}} \right) \approx 0.6 \%
\end{equation}
The direction of tidal stretching is always radial and therefore it's variation shares the orbital frequency. For this reason, if  the tidally induced quadrupole is only 0.6 \% of the orbit induced quadrupole the amplitude of the GW produced by it will only be 0.6 \% of the total amplitude. It is safe to neglect it along with all other smaller contributions.

Now we can proceed with the polarisation prediction. Using the formulae above we find:

\begin{equation}
\begin{split}
    \frac{\Delta L_+(t)}{L}&= \frac{\Delta L_+}{L}  \cos \left(-\frac{L \omega }{c}+2 \omega  t_d\right) \\\
    \frac{\Delta L_\times(t)}{L}&= \frac{\Delta L_\times}{L}  \sin \left(-\frac{L \omega }{c}+2 \omega  t_d\right)
\end{split}
\end{equation}
where $t_d$ is the retarded time and the amplitudes are given by:

\begin{equation}
\begin{split}
   \frac{\Delta L_+}{L}&= \frac{G \mu  R^2 \omega  \cos (2 \phi ) \cos (\psi ) (3+\cos (2 \mathcal{I} )) \sin
   \left(\frac{L \omega }{c}\right)}{c^3 d L}\\\
   \frac{\Delta L_\times}{L}&= -\frac{4 G \mu  R^2 \omega  \sin (2 \phi ) \cos (\psi ) \cos (\mathcal{I} ) \sin
   \left(\frac{L \omega }{c}\right)}{c^3 d L}
   \end{split}
\end{equation}
The ratio $Q$ between the two amplitudes is given by:

\begin{equation}
   Q= \frac{\Delta L_+}{\Delta L_\times}= -  \frac{3 + \cos(2\mathcal{I})}{4\cos(\mathcal{I})} \frac{\cos(2 \phi)}{\sin (2 \phi)}
\end{equation}
Plugging in the values in Table \ref{table:parameters} we find:

\begin{equation}
   \omega = \sqrt{\frac{G\left(m_\text{sdB}+m_\text{WD}\right)}{R^3}} \approx 1.49 \, \text{mHz}
\end{equation}
and:
\begin{equation}
\begin{split}
   \frac{\Delta L_+}{L} &\approx 4.2 \times 10^{-22} \times \cos (2 \phi ) \cos (\psi ) \\\
   \frac{\Delta L_\times}{L} &\approx -1.0 \times 10^{-22}\times \sin (2 \phi ) \cos (\psi ) \\\
   Q&\approx - 4.1 \times \frac{\cos(2 \phi)}{\sin (2 \phi)}
\end{split}
\end{equation}
where the arm length of LISA\footnote{https://www.lisamission.org/articles/lisa-mission/lisa-mission-gravitational-universe} is $L=2.5\times 10^9$ m .
These values can be used to produce the polarisation predictions once the remaining angles are known. Now we proceed to estimate the error and the relative uncertainty in the polarisation amplitudes that is expected from the binary system CD$-30^\circ$ 11223.
We call the uncertainties $\sigma_+$ and $\sigma_{\times}$. By using the values in Table  \ref{table:parameters} and Gaussian error propagation we find:

\begin{equation}
\begin{split}
    \sigma_{+}&\approx 2 \sqrt{ 0.02^2 + \Delta_{\phi}^2 \tan(2 \phi)^2}\\\
    \sigma_{\times} &\approx 2 \sqrt{0.03^2 + \Delta_{\phi}^2\cot(2 \phi)^2}
    \end{split}
    \label{eqn:GW_error_amplitudes}
\end{equation}
\begin{figure}
    \centering
    \includegraphics[scale=0.50]{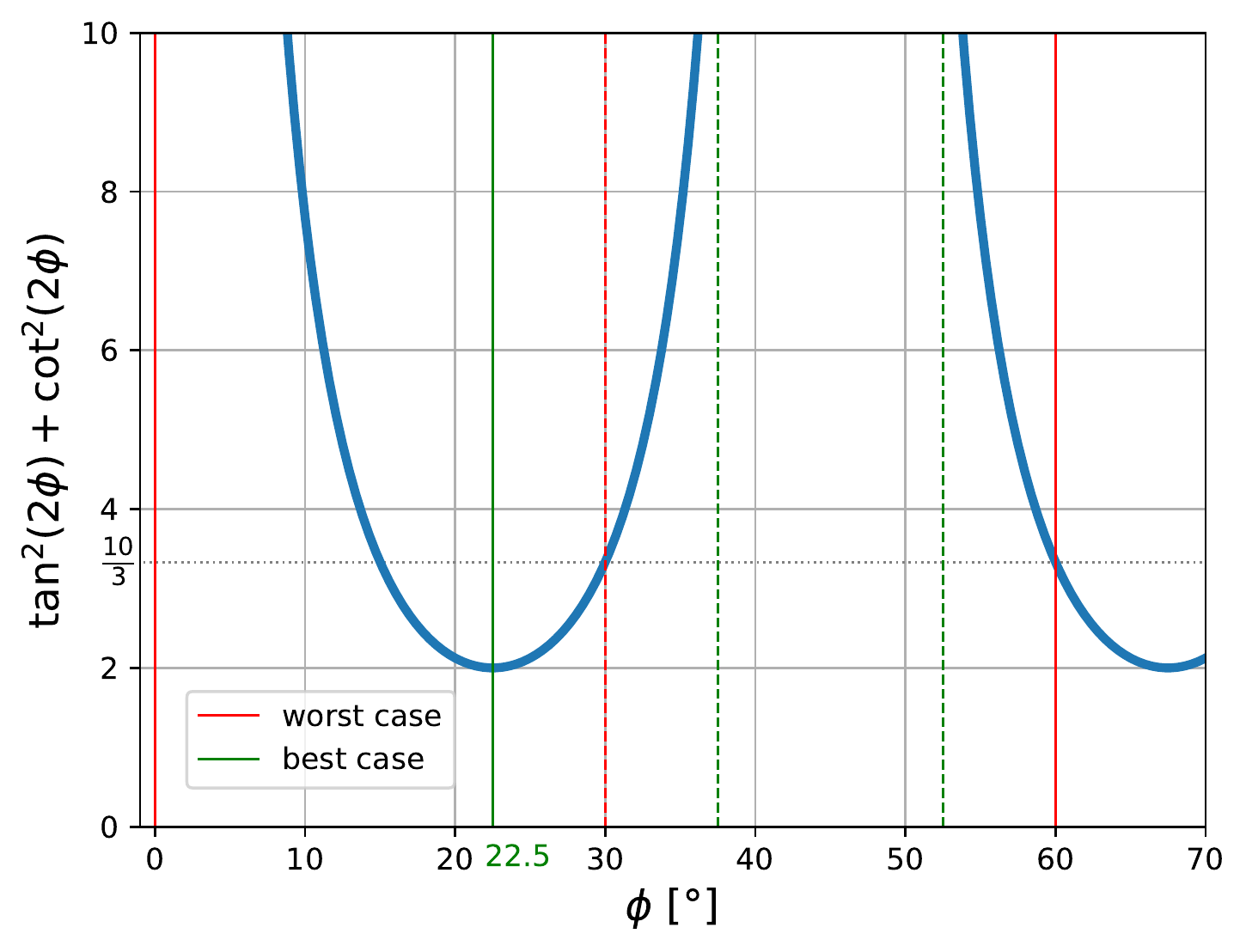}
    \caption{The coefficient $f(\phi)$ in the squared uncertainty $\mathcal{R}^2$ (see Eq. (\ref{eqn:GW_mean_error})) is periodic within 45$^\circ$ ($\pi/4$). The three arms of LISA can sample values of the source's orientation $\phi$ separated by 60$^\circ$ ($\pi/3$). The worst case (indicated by the red vertical lines) in terms of relative error occurs when one arm samples an angle $\phi$ lying on a pole of the function $f(\phi)$. In this case, the measurement with other two arms will correspond to a value $f(\phi)=10/3$. In the best case (indicated by the green vertical lines) one arm samples an angle lying at the minimum of the function $f(\phi)$, such that $f(\phi)=2$. The periodicity of $f(\phi)$ allows to illustrate by what angles the arms would sample in the best ($22.5^\circ$, $82.5^\circ$ and $142.5^\circ$) and worst case ($0.0^\circ$, $60.0^\circ$ and $120.0^\circ$) in less than $180^\circ$. Here we show this by means of the dashed coloured lines.}
    \label{fig:GW_error_plot}
\end{figure}
Here we are considering errors arising from the uncertainties in the parameters $m_{\text{sdB}}$, $m_{\text{WD}}$, $\phi$, $\mathcal{I}$, $R$ and $d$, along with the neglection of the quadrupole moment of the sdB. 
The square sum of the errors $\mathcal{R}^2$ is given by:

\begin{equation}
\begin{split}
    \mathcal{R}^2 = \sigma_{+}^2 + \sigma_{\times}^2 \approx 0.08^2 + 4 \Delta_{\phi}^2 \left(\tan(2\phi)^2 + \cot(2\phi)^2\right)
    \end{split}
    \label{eqn:GW_mean_error}
\end{equation}
The coefficient $f(\phi)=\tan(2\phi)^2 + \cot(2\phi)^2$ plotted in Figure \ref{fig:GW_error_plot} is a periodic function with divergent poles at $\phi = n \pi/4$ and minima at $\phi= \pi/8 + n \pi/4$. Since LISA has three arms approximately resembling an equilateral triangle, it can sample three different values of $\phi$, evenly spaced by sixty degrees, or $\pi/3$. The worst possible case in terms of measurement uncertainties occurs when one of these sampled angles lies on a pole of the function $f(\phi)$. In that case, the other two angles will read $n \pi/4 + \pi/3$ and $n \pi/4 + 2\pi/3$ respectively. The function $f(\phi)$ evaluated at the latter angles yields a value $10/3$. This value represents a ``making the best out of the worst case'' scenario and will be used as a reference in the main text. In reality, the space antennae arms will sample many different orientations as the orbit around the sun. The final formula then reads:

\begin{equation}
\begin{split}
    \mathcal{R}^2 \approx 0.08^2 + \frac{40}{3} \Delta_{\phi}^2
    \end{split}
\end{equation}


\bsp	
\label{lastpage}
\end{document}